\documentclass[useAMS,usenatbib]{mn2e}
\usepackage{psfig}

\newcommand\mdot{\dot{m}}
\def\msun{{\,M_\odot}}

\newcommand\mrad{\dot{m}_{\rm rad}}
\newcommand\msat{\dot{m}_{\rm sat}}
\newcommand\mbh{M_{\rm BH}}
\newcommand\sch{Schwarzschild~}
\newcommand\rozanska{R\'o$\dot{\rm z}$a\'nska }

\newcommand\tvisc{t_{\rm visc}}
\newcommand\fcl{F_{\rm cl}}
\newcommand\fsat{F_{\rm sat}}
\newcommand\sgra{Sgr~A$^*$}

\def\>{$>$}
\def\<{$<$}

\def\simlt{\lower.5ex\hbox{$\; \buildrel < \over \sim \;$}}
\def\simgt{\lower.5ex\hbox{$\; \buildrel > \over \sim \;$}}
\def\sqr#1#2{{\vcenter{\hrule height.#2pt
      \hbox{\vrule width.#2pt height#1pt \kern#1pt
         \vrule width.#2pt}
      \hrule height.#2pt}}}

\def\del#1{{}}

\title[Condensing accretion flows]{Feeding the black hole with
condensing accretion flows: radiatively efficient and radiatively
inefficient cases.}

\author[Nayakshin]{Sergei Nayakshin\\ Max-Planck-Institut f\"ur
Astrophysik, Karl-Schwarzschild-Str. 1, 85740 Garching, Germany}
\date{2003 Xxxxx XX}

\pagerange{\pageref{firstpage}--\pageref{lastpage}} \pubyear{2003}

\def\LaTeX{L\kern-.36em\raise.3ex\hbox{a}\kern-.15em
    T\kern-.1667em\lower.7ex\hbox{E}\kern-.125emX}

\begin{document}

\label{firstpage}

\maketitle

\begin{abstract}
We study the accretion flow of a hot gas captured by the black hole
gravity in the presence of a thin cold accretion disk. Such
geometrical arrangement is expected in Active Galactic Nuclei (AGN)
and in galactic X-ray binary systems because both hot and cold gases
are present in the black hole vicinity. Previous astrophysical
literature concentrated on the evaporation of the cold disk in the
classical heat conduction limit. Here we consider the inverse process,
i.e. condensation of the hot gas onto the cold disk. We find two
distinct condensation regimes. (i) In the classical thermal conduction
limit, the radiative cooling in the hot gas itself force condensation
above a certain critical accretion rate. Most of the flow energy in
this case is re-emitted as X-ray radiation.  (ii) Below a certain
minimum accretion rate, the hot electrons are collisionless and the
classical heat flux description becomes invalid.  We use the
``non-local'' heat flux approach borrowed from the terrestrial laser
heated plasma experiments. Due to their very large mean free path, the
hot particles penetrate deep into the cold disk where the radiative
losses are significant enough to enable condensation. In this case the
hot flow energy is inconspicuously re-radiated by the transition layer
in many UV and especially optical recombination lines (e.g.,
Ly$\alpha$, $H\alpha$, H$\beta$) as well as via the optically thick
disk emission.  We derive an approximate analytical solution for the
dynamics of the hot condensing flow. If the cold disk is inactive,
i.e.  accumulating mass for a future accretion outburst, then the
two-phase flows appear radiatively inefficient. These condensing
solutions may be relevant to \sgra, low luminosity AGN, and transient
binary accreting systems in quiescence.
\end{abstract}

\section{Introduction}

An optically thick accretion disk (e.g., Shakura \& Sunyaev 1973)
appears to be an important part of the accretion flow of gas into the
black hole (BH) or a compact object. This conclusion is supported by
spectral energy distributions (SED; see Elvis et al. 1994; Ho 1999;
and Fig. 1 in Gierlinski et al. 1999), double peaked emission line
profiles (e.g. McClintock et al. 2003; and references in Ho 2003), and
eclipse mapping of binary systems (e.g. Wood et al. 1986). Even for
\sgra, a very dim source, there is now a suspicion that a cold {\em
inactive}, i.e. not accreting, disk may be present (Nayakshin, Cuadra
\& Sunyaev 2004).

There is also a significant amount of hot ($T > 10^7$ K) gas at large
distances from the BH in galactic nuclei. In the best studied cases
this hot gas is observed as close as its capture radius (e.g. in the
giant eliptical galaxy and a LLAGN M87 [Di Matteo et al. 2003]; and in
\sgra\ [Baganoff et al. 2003]), meaning that accretion of this hot gas
on the BH is unavoidable. In galactic binary systems, a diffuse hot
gas may be present near the disk outer rim due to a shock in the hot
spot, if the radiative cooling time is longer than dynamical time. A
shocked captured stellar wind from the secondary is another source of
hot gas (and it may also form the cold disk itself; Kolykhalov \&
Sunyaev 1980).

Thus the accretion flow at large distances is often a two-flow problem
(see Figure \ref{fig:geometry}).  Thermal conduction in a multi-phase
medium leads to either evaporation of the cold gas, or vice versa,
condensation of the hot gas (e.g., Zel'dovich \& Pickel'ner 1969;
Penston \& Brown 1970).  In reference to accretion flows, Meyer \&
Meyer-Hofmeister (1994; MMH94 hereafter) were the first to study the
mass exchange between the cold disk and the corona. Their pioneering
study, extended later by, e.g., Liu, Meyer \& Meyer-Hofmeister (1997);
Dullemond (1999); \rozanska \& Czerny (2000a,b), showed that the cold
disk evaporates at low coronal accretion rates. At large accretion
rates the radiative cooling suppresses the corona (\rozanska \& Czerny
2000a). The magnitude of the viscosity coefficient, $\alpha$ (Shakura
\& Sunyaev 1973), was shown to be extremely important. In all the
models the evaporation rate decreases very strongly with decreasing
$\alpha$ (e.g.  Meyer-Hofmeister \& Meyer 2001). Recently, Liu, Meyer
\& Meyer-Hofmeister (2004) found condencing solutions for a
sufficiently small value of the viscosity parameter.

Here we aim to study the condensation process systematically,
i.e. attempting to cover a broad range of accretion rates in the hot
flow. We first review the classical thermal conduction case. As
expected, we recover qualitatively the results previously obtained by
the above referenced authors. Namely, when the heat flux is classical,
we find condensation at high and evaporation at low coronal accretion
rates (\S \ref{sec:clas}). The critical accretion rate (at which no
mass exchange between the disk and the hot corona takes place) is a
strong function of $\alpha$ (\S\S \ref{sec:clascond} \&
\ref{sec:paramspace}).

We then point out that for every value of $\alpha$, there exists an
accretion rate below which the classical thermal conduction treatment
becomes invalid since the hot electron mean free path, $\lambda$, is
long compared with the flow height scale, $H$. To study the problem in
this limit, we use a modified (see \S\S \, \ref{sec:basic} \&
\ref{sec:vertical}) prescription for the heat flux that is borrowed
from the laser-heated plasma experiments.  The heat flux is then
proportional to the saturated heat flux coefficient, $\phi\le
1$. While this case still requires a future physical kinetics
treatment, we show that there may exist an additional condensation
regime. In this regime the hot particles penetrate deep into the cold
gas. The cold gas density turns out to be great enough to re-radiate
the deposited energy away, enabling condensation. With the help of
certain approximations, we build an analytical solution (\S\S\,
\ref{sec:dynamics} \& \ref{sec:analytical}) for the radial structure
of the hot flow in this case.

We put our results into the context of the previous work in \S
\ref{sec:paramspace}, where we determine the type of solution for a
given combination of parameters (accretion rate, radius, $\alpha$ and
$\phi$). We discuss the expected spectra from the two types of the
condensing flows in \S \ref{sec:obs}. The classical thermal conduction
condensing flow emits mainly in X-rays. Such solutions apply at
relatively high accretion rates, and therefore they are likely to be
relevant to bright or medium-bright AGN, such as Seyfert Galaxies. Due
to these high accretion/condesation rates, the cool disk is probably
active, meaning that the SED of these sources would be dominated by
the emission from the small scale accretion disk in the vicinity of
the last stable orbit. Such flows would obviously be radiatively
efficient.

The second (non-local) condensation regime is applicable at lower
accretion rates. It is therefore possible that accretion disks fed by
condensation at these low rates would be inactive, as in quiescent
states of transient binary sources. In this case the hot gas, settling
onto the cold disk, effectively stops accreting. The hot accretion
flow is thus terminated at some large distance away from the central
object, which implies that the radiation emitted by such a flow will
be much less luminous than expected from a flow onto the black
hole. Furthermore, these flows may be quite dim in X-rays since the
raditive cooling in the corona itself is small. The cold disk serves
as a huge cooling plate (radiator) for the hot flow. The SED of such
flows would then be dominated by their infra-red/optical thermal-like
bumps rather than UV or X-ray frequency regions. We suggest that this
radiative ``inefficiency'' (rather a time-delayed accretion) of
condensing flows may be one of the reason why some LLAGN, and the
galactic black hole \sgra, appear underluminous.

\section{Classical thermal conduction in two-phase flows}\label{sec:clas}

\subsection{Basic physics of thermal conduction}\label{sec:basic}

\begin{figure}
\centerline{\psfig{file=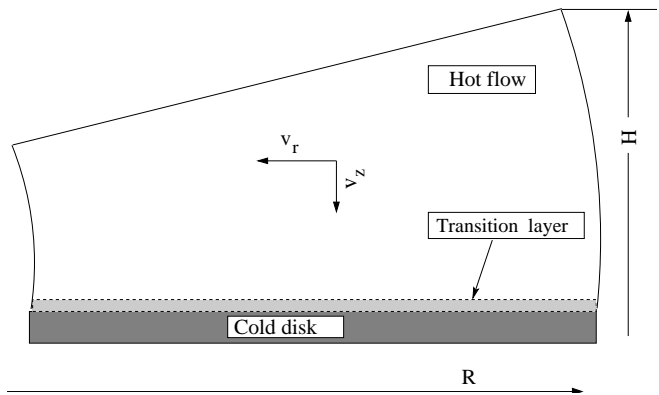,width=.49\textwidth,angle=0}}
\caption{Geometry of the problem (shown for $z>0$ only). The hot flow
(corona) sandwiches a cold accretion disk. A transition layer heated
by hot coronal electrons (\S \ref{sec:vertical}) develops on the top
of the cold disk.  In the limit of collisionless thermal conduction,
both the transition layer and the disk are thick to coronal electrons,
and are very geometrically thin. In contrast, the corona is thin to
coronal electrons (since $H \simlt \lambda$), but is geometrically
thick. Depending on conditions, the hot flow may be condensing
(positive $v_z$) or evaporating. In the classical thermal conduction
limit, the transition layer becomes optically thick to hot
electrons. The layer's geometrical thickness also increases.}
\label{fig:geometry}
\end{figure}

Figure \ref{fig:geometry} shows a sketch of geometry that we envisage
in this paper. The hot flow is sandwiching a very cold and thin disk
located in the midplane. A thin (compared with $R$) transition layer
develops between the corona\footnote{Throughout the paper, we shall
use the terms ``corona'' and the ``hot flow'' interchangeably,
although we do not assume that the origin of the corona is the cold
disk.} and the disk. Since the cold disk viscous time is very long,
the disk can be treated as stationary.

The physics of the mass exchange between the two flows depends greatly
on the heat flow from the hot to the cold gas.  When collisions are
abundant, i.e the mean free path for electron energy exchange,
$\lambda$, is smaller than the temperature scale height, $L_T =
T/|\nabla T|$, the classical (Spitzer \& Harm 1953; Spitzer 1962) heat
conductivity is used. The thermal heat flux, $F$, in this case is
given by
\begin{equation}
\vec{F} = \vec{F}_{\rm cl} = - k(T) \nabla T \;,
\label{spitzer}
\end{equation}
where $T$ is the temperature (we will assume that electron temperature
is equal to the proton's temperature), $k(T)$ is the conductivity
coefficient given by
\begin{equation}
k(T) = 1.84 \times 10^{-5} T^{5/2}/ \ln \Lambda_c \quad\hbox{erg
cm$^{-1}$ s$^{-1}$ K$^{-1}$}\;,
\label{k}
\end{equation}
where $\ln \Lambda_c\sim 30$ is the Coulomb logarithm. The heating or
cooling rate due to thermal conduction is
\begin{equation}
H_{\rm cl} = -\, \hbox{div}\,\vec{F}\;.
\label{hcl}
\end{equation}
For reference, 
\begin{equation}
\lambda\simeq 10^4 T^2/n\; \hbox{ cm}
\label{lcm77}
\end{equation}
(Spitzer 1962), where $n$ is the gas density in cm$^{-3}$.  If
collisions are rare, i.e. $\lambda > L_T$, the classical formula
over-estimates the heat flux. The electron conduction is then
artificially limited to (e.g. Parker 1963; Cowie \& McKee 1977) the
saturated heat flux:
\begin{equation}
\fsat = 5 \phi \rho c_s^3\;,
\label{fsat}
\end{equation}
where $P$ and $\rho$ are the gas pressure and density, respectively,
$c_s^2= P/\rho$ is the local isothermal sound speed of the gas, and
$\phi\le 1$ is the saturation coefficient. Even though magnetic fields
may be important in reducing the thermal conductivity (see Appendix),
in what follows their effects are included only through the magnitude
of $\phi$. A more detailed discussion of thermal conductivity in the
saturated regime is given in \S \ref{sec:vertical}.

\subsection{Classical evaporation or condensation}\label{sec:clascond}

Numerically accurate global evaporating solutions were obtained by
MMH94, and that for condensation by \rozanska \& Czerny (2000a).  Here
we only wish to establish the sign of the mass exchange between the
two flows for given {\em local} conditions in the hot gas.  For the
sake of simplicity and generality, we approximate the geometry
(Fig. \ref{fig:geometry}) as planar, and neglect possible winds. In
this setup the problem is similar to that of an infinite
plane-parallel two-phase system studied by Zel'dovich \& Pikel'ner
(1969), and Penston \& Brown (1970). In such a system, for a subsonic
flow, the entropy equation (for adiabatic index $\gamma=5/3$ and for
$v_z>0$ for downward direction; Fig. \ref{fig:geometry}) is
\begin{equation}
- \rho v_z \frac{5 k_B} {2 \mu} \frac{d T}{d z} +
\frac{d F}{dz} = 
q_+ - \Lambda(T) n^2 \;,
\label{est}
\end{equation}
where $z$ is the vertical coordinate, $q_+$ is the heating rate in
erg/cm$^3$/sec, $\Lambda(T)$ is the usual optically thin cooling
function, and $k_B$ and $\mu$ are the Bolzmann's constant and the
mean molecular weight in the corona. Note that the subsonic flow
assumption can be verified a posteriorly: as results of Cowie \& McKee
(1977) for evaporation of spherical clouds show, the flow is always
subsonic when the heat flux is classical. Also note that $|v_z| \ll
c_s$ implies that the pressure is approximately constant throughout
the thermal conduction interface.

The boundary conditions for
this equation are $T(z=0)=T_c$, $F(0)=0$, and that the heating is
equal to cooling: $q_+(0) - \Lambda(T_c) n^2(0)=0$. Analogously,
$T(\infty)=T_h$, etc.

Multiplication of equation \ref{est} by the heat flux and integration
from $z=0$ to $z = \infty$ yields
\begin{equation}
\rho v_z\; = - \frac{I_1}{I_2}\;,
\label{j}
\end{equation}
where $I_1$ and $I_2$ are the integrals 
\begin{equation}
I_1 = \int_{T_c}^{T_h} dT\; T^{5/2} \left[ \;q_+\; -
\;\Lambda(T)n^2(z)\;\right]\;,
\label{i1}
\end{equation}
and 
\begin{equation}
I_2 = \frac{5k_B}{2\mu}\;\int_{0}^{\infty} dz\; T^{5/2}
\left(\frac{dT}{dz}\right)^2\;,
\label{i2}
\end{equation}
respectively. Now, note that $I_2$ is always positive, and hence the
sign of the mass exchange depends only on that of the integral
$I_1$. Further, the cooling rate is proportional to $\rho^2$ while the
heating rate often is proportional to just $\rho$, thus for coronal
density $\rho_h$ high enough, the radiative cooling forces $I_1<0$,
and the hot flow will condense. Conversely, for low enough $\rho_h$,
the integral is positive and hence the evaporation is taking place.
Evidently, for a given value of $T_h$ a unique value of pressure, at
which the mass exchange is zero, may be defined. This pressure has
been termed the ``saturated vapor pressure'' by Penston \& Brown
(1970).

Consider now in more detail integral $I_1$. Above the peak at $T\simeq
10^5$ in the cooling curve, where $\Lambda(T)\simeq 10^{-21}$ erg
cm$^3$ sec$^{-1}$, following McKee \& Cowie (1977), we approximate the
radiative cooling function as $\Lambda(T)\propto T^{-0.6}$ (until the
free-free radiation losses become important). Since at a constant
pressure $n\propto T^{-1}$, $\Lambda(T) n^2 \propto T^{-0.6} T^{-2}
\propto T^{-2.6}$, one could expect that the atomic cooling peak would
significantly contribute to the integral. However, the $T^{5/2}$
factor strongly favors high temperatures, and as a result we have
$\int dT \Lambda(T) (n T)^2 T^{3/2} \propto p^2 \int dT \Lambda(T)
T^{1/2} \propto \int dT \Lambda(T) T^{1/2} \propto T^{0.9}$. This
shows that the cooling contribution to $I_1$ is strongly dominated by
the highest temperatures, i.e. by $T\simlt T_h$. The heating rate is
usually a weaker function of temperature and density than the cooling
rate, thus the latter statement is even more appropriate. Therefore,
approximately, we have
\begin{equation}
I_1 \sim T_h^{7/2} \left[ \;\overline{q}_+\; -
\;\overline{\Lambda(T)n^2(z)}\;\right]\; \propto \; Q_+ - Q_-\;,
\label{i1ap}
\end{equation}
where the vertical bar signifies averaging over the hot flow, and the
symbols $Q_+$ and $Q_-$ are the vertically integrated heating and
cooling rates in the corona region.

We shall recall that for typical heating and cooling mechanisms, a
two-phase system without a thermal conduction may be thermally
unstable over a certain range in pressure (e.g. Field 1965; Krolik,
McKee \& Tarter 1983). In particular, an initially uniform hot one
phase medium will become thermally unstable and develop condensations
when a critical pressure value is exceeded. Now, an important point is
that this critical pressure and the saturated vapor pressure are
usually rather close. For example, in the common
Compton-bremsstrahlung case (McKee \& Begelman 1990), the critical
pressure is 0.95\% of the pressure at which thermal instability in the
hot gas starts to form clumps. The foregoing discussion and equation
\ref{i1ap} explain this result. As the integral $I_1$ is dominated by
temperatures $T\simeq T_h$, thermal conduction drives condensation at
approximately same conditions as those when the thermal instability
starts to form condensations in the hot flow.

In the context of an accretion flow, the critical pressure or density
can be used to define a critical accretion rate, $\dot{m}_{\rm crit}$.
In this paper we measure accretion rates in units of the Eddington
accretion rate, defined as $L_{\rm Edd}/0.1 c^2$, where $L_{\rm Edd} =
1.4 \times 10^{46} M_8$ erg s$^{-1}$, and $M_8$ is the black hole
mass, $M_{\rm BH}$, in $10^8$ Solar masses.  In the Non-Radiative
Accretion Flow theories (NRAFs, e.g. Narayan \& Yi 1994; Blandford \&
Begelman 1999, BB99 hereafter), the radiative cooling term is not
important ($Q_-\simeq 0$) as the gas is very tenuous. However, as the
accretion rate increases, gas density increases, the radiative loss
term eventually becomes important, and a radiative collapse of the hot
(one-phase) flow occurs when $Q_- \simeq Q_+$.  The corresponding
critical accretion rate, $\dot{m}_{\rm crit} \sim \alpha^2 r_4^{-1/2}$
(e.g. Narayan \& Yi 1995; Esin 1997), where $r_4$ is radius in units
of $10^4 R_S$, with $R_S = 2 GM_{\rm BH}/c^2$, the \sch radius.  The
arguments made above for the {\em classical} thermal conduction show
that the critical accretion rate at which the mass exchange in the
two-phase flow is zero, $ \dot{m}_{\rm clas}$, is quite close to
$\dot{m}_{\rm crit}$.  In a simple model for the two-phase flow with
$q_+ = 9/2 \alpha \rho c_s^2 \Omega_K (1 - T/T_{\rm vir})$, where the
$(1-T/T_{\rm vir})$ approximately accounts for the effects of winds,
and free-free cooling, we found that the conduction drove evaporation
at $\mdot_{\rm clas}\approx 0.7 \mdot_{\rm crit}$.

Summarizing these results, within the classical heat flux
approximation, the hot flow evaporates the cold disk for accretion
rates smaller than $\mdot_{\rm clas}$, and condenses for $\mdot >
\mdot_{\rm clas}$. The classical condensation accretion rate
$\mdot_{\rm clas}$ is only slightly lower than the critical accretion
rate $\mdot_{\rm crit}$ at which the hot non-radiative flows without
underlying disks would collapse to a thin disk. In the NRAF theories
the radiative cooling term is not important, and hence these flows
should drive an evaporation of the cold disk.

\del{However, in a classical thermal conduction interface, the column
depth of such cold material is very small, and this is why it cannot
contribute to the integral $I_0$ significantly. Another useful way to
view this result is to note that because of the strong temperature
dependence of the classical heat flux (equation \ref{k}), very little
of the heat flux is able to propagate down into the cool layers of the
gas.}

\section{Condensation/evaporation in collisionless
limit}\label{sec:vertical} 

\subsection{The non-local heat flux}\label{sec:nonlocal}

As accretion rate decreases, density decreases in non-radiative
accretion flows. Thus at some point $\lambda$ (equation \ref{lcm77})
will become larger than the scale height $H < R$ of the hot flow, and
the classical heat flux description will become invalid (the exact
condition for this is discussed in \S \ref{sec:sar}). The common
approach in the astrophysical literature is to limit the heat flux to
the saturated heat flux value, as described in \S
\ref{sec:basic}. However, although the saturated heat flux
prescription remedies the situation as far as the maximum heat flow is
concerned, it appears to be unreliable in regard to the correct
estimate of the conductive heating rate.  Namely, the latter is
calculated as $d\fsat/dz = 5 \phi P d c_s/dz = 5 \phi P (c_s/2T)
dT/dz$. While the heat flux is finite in the limit $dT/dz \rightarrow
\infty$, the thermal conduction heating is infinite. Physically, the
maximum heating rate should be given by the heating rate of a cold
test particle immersed in the hot gas (in fact it is far lower than
that due to self-generated electric fields).

A detailed calculation of the vertical structure of the transition
region between the hot and the cold gas is beyond the scope of the
present paper (but will be taken up in Nayakshin \& Sunyaev 2004, in
preparation; NS04 hereafter). The problem should be clearly treated
with a physical kinetic approach rather than a hydrodynamical
one. However we can take guidance from the laser-heated plasma (LHP)
research, in which the thermal conduction in the collisionless
(saturated) regime has been thoroughly studied via direct physical
kinetics calculations and {\em in-situ} experimental measurements. At
a physical level, the LHP research showed the importance of the high
energy electrons that have very long mean free paths. These electrons
carry most of the flux and their distribution function is not coupled
to the the average quantities such as the mean temperature.

Luciani, Mora \& Virmont (1983; LMV hereafter), and Luciani, Mora \&
Pellat (1985) were able to formulate a ``non-local'' heat flux
prescription that operates with {\em hydrodynamical} not physical
kinetics quantities, and is yet able to capture the majority of the
relevant physics. The price that one has to pay for a greater
self-consistency is that the heat flux becomes a non-local
quantity. In particular, LMV expressed the heat flux, $F_{\rm nl}$, as
an integral -- a convolution of the classical Spitzer's flux with a
kernel that has a width of $\sim \lambda$:
\begin{equation}
F_{\rm nl}(z) = \int_{-\infty}^{+\infty} \frac{dz'}{2\lambda(z')}\;
\fcl(z') \exp\left[-\tau(z,z')\right]\;,
\label{fnl}
\end{equation}
where $\tau(z,z')$ is the ``electron optical depth'' between location
$z$ and $z'$,
\begin{equation}
\tau(z,z') = \left|\frac{1}{\lambda(z')n(z')}\; \int_{z}^{z'} dz''
n(z'')\right|\;.
\label{tauel}
\end{equation}
(Note that $\lambda$ used by LMV is about a factor of 2 larger than
the electron equipartition $\lambda$ mentioned in \S \ref{sec:basic};
NS04). Formula \ref{fnl} has a simple physical interpretation: as the
heat flux is dominated by electrons with energy $E\simgt 5 k_B T$
(e.g. see \S IIA in Schurtz et al. 2000 for a nice account of this
fact), their behavior is non-local. They are practically free
streaming and thus the exponential factor in equation \ref{fnl} is an
attempt to model their transfer.

Formula \ref{fnl} reduces to the classical limit when $\lambda\ll
L_T$, as expected. In addition, the non-local heat flux may be shown
to scale as the saturated heat flux\footnote{The normalizing factor
$\phi$ in equation \ref{fsat} models complicated plasma collective
effects, magnetic fields, etc., and may also be introduced in
equation \ref{fnl}.} at an infinitely sharp temperature discontinuity,
as it should.  Most importantly, the heating rate due to the non-local
electron conduction -- the derivative of $F_{\rm nl}$ (equation
\ref{hcl}) -- acts now on the kernel underneath the integral sign and
hence it is finite.

Although semi-empirical, formula \ref{fnl} has been checked against
numerical Fokker-Planck simulations (starting from Luciani et
al. 1985). There have been improvements to the LMV kernel, and new
non-local kernels were suggested (e.g. Albritton et al. 1986; also
Epperlein \& Short 1991, and references there). However they are all
based on the same physical ideas, and the LMV approach still remains
the most popular choice (e.g. Ditmire et al. 1998; Schurtz, Nicolai \&
Busquet 2000) perhaps due to its relative simplicity.

\subsection{Vertical temperature profile}\label{sec:temp}

\begin{figure}
\centerline{\psfig{file=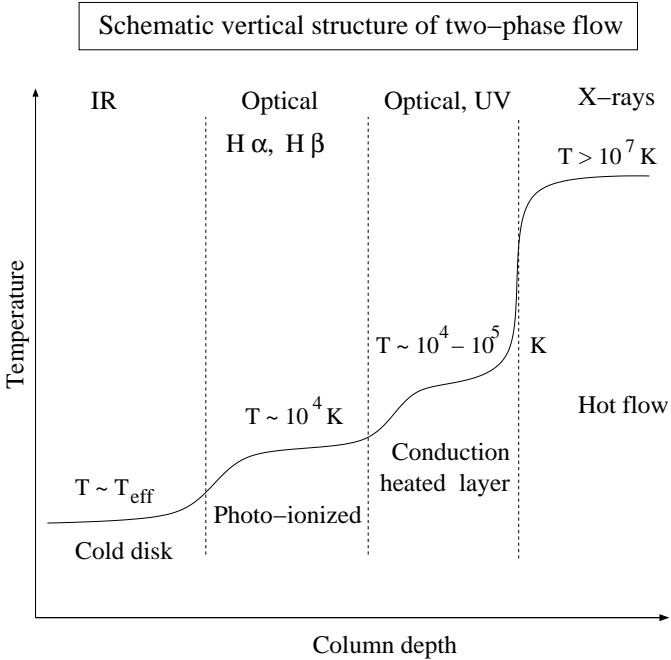,width=.5\textwidth,angle=0}}
\caption{Schematic vertical temperature profile of the two-phase flows
(not to scale).  The ``cold disk'' region is the bulk part of the
disk. Directly above it is the layer heated and photo-ionized by
emission of the gas above it (X-rays and possibly UV). The $T\sim
10^4- 10^5$ K region is conduction-heated by the energetic electrons
penetrating directly from the hot flow. The depth of this layer is of
order one mean free path of hot electrons.}
\label{fig:temp-sketch}
\end{figure}

The expected vertical temperature structure of the two-phase flow with
a transition region is shown schematically in Figure
\ref{fig:temp-sketch}.  The $x$-axis of the plot shows the column
depth from the disk midplane, i.e. $N=\int_0^z dz' n(z')$. The corona
is situated on the top and its temperature, $T_h \ge 10^7$ K (may be
much hotter, say $10^9$ K). At the lower boundary of the hot flow, gas
``temperature'' rapidly decreases over a thin layer separating the
corona from the ``conduction heated zone''. In the both the classical
and the saturated heat flux theories, almost all of the heat carried
by the hot electrons should be dissipated inside the region of the
high temperature gradient. As the radiative losses there are small,
this heating would be unbalanced and would lead to a vigorous
evaporation. However, both the classical and the saturated heat flux
formulations fail to describe this region properly. The particle
distribution function in the region is strongly non-Maxwellian as it
consists of a mix of the hot and the cold particles. While the local
``temperature'' does define the mean particle kinetic energy, it is
not related to the energy dispersion at all. The non-local heat flux
predicts that only a small fraction of the heat flux, i.e. $\Delta
z/\lambda$, where $\Delta z$ is the width of the transition zone, will
be liberated in the region of the steep temperature gradient. Thus the
evaporation rate of this region is strongly reduced compared with the
saturated heat flux predictions (NS04).

\del{The evaporation rate can be
estimated as 
\begin{equation}
\mdot_{+} \simeq \frac{\Delta \fsat}{5 k_b T_h/2 \mu} 
\label{mplus}
\end{equation}}

The rest of the non-local heat flux enters the conduction-heated
transition layer, in which the particle distribution is dominated by
the cold gas, and whose temperature is $\simlt 10^5$ K as this
corresponds to the peak of the optically thin cooling curve. Hydrogen
in the layer is collisionally ionized. Below we shall show that the
layer is dense enough to re-radiate away the deposited thermal
conduction heating. The emission occurs mostly in the optical and UV
frequencies. The column depth of the conduction-heated zone is about
one penetration depth, i.e. $N_p = \lambda n \simeq 10^4 T_h^2
\hbox{cm}^2$. Since the hot flow by definition is optically thin to
the hot electrons, the transition layer is actually optically thicker
than the latter.

Roughly a half of the radiation emitted by the conduction-heated zone,
plus X-rays from the corona, is directed to and penetrate into a
deeper cooler layer of the disk, photo-ionizing hydrogen there. This
deeper layer is thus essentially an HII region. Finally, yet deeper is
the main body of the disk. The disk is assumed to be optically thick
and be at least as hot as the effective temperature corresponding to
the heat flux from the hot flow ($\sigma_B T_{\rm eff}^4 \simeq
\fsat$).

\subsection{Energy balance in the conduction-heated
zone}\label{sec:radzone}

The heating rate due to the non-local electrons penetrating the cold
gas from the corona is
\begin{equation}
H_{\rm nl} = \left|\frac{d F_{\rm nl}}{dz}\right| =
\frac{\fsat}{\lambda_{\rm hc}} \exp\left[-\tau(z)\right] = \frac{5
\phi P c_s}{\lambda_{\rm hc}} \exp\left[-\tau(z)\right]\;,
\label{hnl}
\end{equation}
where we set the maximum of the non-local heat flux to equal to the
saturated heat flux in the corona, at it should; $\tau(z)$ is the
optical depth of the hot electrons measured from the bottom of the
corona, $\lambda_{\rm hc}$ is the mean free path of hot electrons {\em
in cold gas}, $\lambda_{\rm hc} = N_p/n_c$, and $n_c$ is the cold
(that is local) gas number density. Note that the heating rate is
simply given by the heat flux ($\fsat$), divided by the effective
distance to which the hot particles penetrate ($\lambda_{\rm hc}$),
plus the attenuation factor ($\exp(-\tau(z)$).

Note that everywhere except the hot corona the pressure scale height
is much larger than the temperature scale height, and hence the
pressure is roughly constant with height (until the top coronal
layers). With this, we have numerically
\begin{equation}
H_{\rm nl} \simlt 10^{-22} \, \phi \, x_e \, T_5 T_8^{-3/2} n_c^2\;,
\label{hcomp}
\end{equation}
where $x_e\le 1$ is the fraction of free electrons in the transition
layer, $T_5 = T_{\rm c}/10^5$ K is its temperature, and $T_8 =
T_h/10^8$ K is the hot flow temperature. Introduction of $x_e$ above
is a rudimentary attempt to account for the change in the energy
exchange rate, and is approximately valid as long as $x_e \simgt [\ln
\Lambda_c]^{-1}$. At lower $x_e$ ionization of atomic Hydrogen, rather
than Coulomb collisions, becomes the dominant heating mechanism.

The cooling rate of the transition layer is $\Lambda(T_c) n_c^2$. The
maximum of the cooling function for collisionally ionized gas is
around $\Lambda_{\rm max} \sim 10^{-21}$ erg/sec cm$^3$ at temperature
$T_c \simeq 10^5$ K (e.g. Raymond, Cox \& Smith 1976). Comparing this
with the heating rate given by equation \ref{hcomp}, we observe that
this heating {\em can} be compensated by the atomic cooling if $T_h
\simgt \hbox{few}\times 10^7$ K or so. In addition, the
non-equilibrium ionization state of the gas can further increase the
cooling function by a factor of up to few tens (e.g., see Fig. 5 in
Borkowski et al. 1990). This (i.e. $\Lambda(T) n^2 > H_{\rm nl}$) is
the regime that we are interested in.

It is instructive to compare the cooling rates in the hot corona and
in the conduction-heated layer. The former is $\sim 10^{-23} n_h^2$
erg cm$^3$/sec, whereas the latter is $\sim 10^{-21} n_c^2 = 10^{-21}
(T_h/T_c)^2 n_h^2$, i.e. typically some 6-8 orders of magnitude
larger. This is why the cold disk is such an efficient coolant for the
hot coronal flow.

\subsection{Condensation of hot gas}\label{sec:ions}

Up till now we have considered thermal conduction due to electrons
only. The ion thermal conductivity is neglected in all three
formulation -- the classical, saturated and the non-local -- for the
reason that ions are much heavier and thus much slower than the
electrons, thus carrying much less flux. However, in the saturated
heat flux conditions, the electron heat flow is limited by factors of
many to few tens compared with their free-streaming flux, most likely
due to charge conservation and plasma instabilities arising when
electrons stream past the ions. Thus the ion heat flow may become more
important.

Coulomb heating rates of cold gas by hot Maxwellian electrons and ions
are well known (Spitzer 1962). In the problem of interest, the
situation is far more complex since it is not well known whether
and how the non-thermal ions will stream into the cold gas (e.g. see
Balbus \& McKee 1982). Below we shall make the simplest assumption,
namely that the ratio of the non-local ion heating rate of the cold
gas to that done by the non-local electrons (equation \ref{hnl}) is
the same as that for a Maxwellian distribution. Thus the total ion
(assumed to be proton) and electron non-local heating rates are 
\begin{equation}
H_{\rm tot} = H_{\rm nl}\; G(T, T_h)\;,
\label{htot}
\end{equation}
where the factor $G$ is (Spitzer 1962; Balbus \& McKee 1982):
\begin{equation}
G = 1 + \frac{43 \; f_p}{\left(1 + 1836\; T/T_h\right)^{3/2}}\;.
\label{g}
\end{equation}
The factor $f_p < 1$ is introduced to parameterize the fraction of the
heat flux carried by the ions.  Note that $G\rightarrow 1$ for $T \gg
T_h/1000$ and that $G \rightarrow 1 + 43 f_p$ for $T \ll T_h/1000$.
As an example, for $f_p=0.5$, $T=10^5$, and $T_h = 3\times 10^7$ K,
$G\simeq 2.2$, whereas for $T_h = 10^8$, $G=5.5$.

Now, if the fraction of the non-local flux dissipated in the
transition layer between the conduction-heated zone and the corona is
negligible, and {\em if} $\Lambda(T) n^2 > H_{\rm tot}$, condensation
will prevail over evaporation. Since $G$ is not much larger than unity
for $T_h\simlt 10^8$ K, and since for larger $T_h$ the total heating
rate decreases as $T_h^{-3/2}$ (equation \ref{hcomp}), it appears that
the condensation will be taking place when the corona temperature is
larger than the ``condensation temperature'' $T_{\rm cond} \sim 10^7$
K.

It is likely that the hot flow temperature at the outer radius $R_0$
(see \S \ref{sec:rad}) is nearly virial. The latter exceeds $T_{\rm
cond}$ for radii smaller than $\sim 10^5 R_S$, i.e. in most
astrophysically interesting situations (e.g. cf. Fig. 5 in Narayan
2002). Also note that X-ray spectra of LLAGN are usually power-laws in
2-10 keV energy band, and thus on purely observational grounds the
temperature of hot flows should be larger than $10^7$ K for these.

\del{condensation rate of the hot
ions is
\begin{equation}
\mdot_- = \frac{f_p \fsat}{E_p} \equiv \; {\cal M}_c\, \rho_h c_h
\label{mmin}
\end{equation}
where $E_p \sim \hbox{few}\; k_B T_h$ is the average kinetic energy of
the hot ions entering the cold disk, and $\rho_h$ and $c_h$ are the
average gas density and ion sound speed in the corona. ${\cal M}_c$ is
the Mach number of the condensing flow, ${\cal M}_c = 2 \phi f_p (5
k_B T_h/2 E_p) < 1$.}

\section{Radial dynamical equations for hot flow}\label{sec:dynamics}

We now wish to write down simplified equations for the hot condensing
flow with a goal of providing an example of the radial structure of
such flows. We shall do this only for the newly found non-local
condensation regime (\S \ref{sec:vertical}), when the heat flux is
saturated.

\del{Due to the fact that depending on parameters, condensation may be
proceeding via either radiative (\S \ref{sec:clas}) or the non-local
limit,}

\subsection{Domain of applicability in radius}\label{sec:rad}

We study the problem at radii smaller than $R = R_c$ where $R_c$ is
the circularization radius of the hot gas. For one-phase accretion
flows, the radial inflow velocity (e.g. see \S 5 Frank, King \& Raine
2002) of the hot flow is $v_R \sim \alpha c_s (H/R)^2 \sim \alpha c_s$
(for $H/R\sim 1$). Using this we obtain an estimate of the coronal gas
density $n$ at any radius $R$ and a given accretion rate. We then
compare the hot electron mean free path with $R$:
\begin{equation}
\frac{\lambda}{R}\sim 0.1\; \alpha\, \mdot^{-1} r_4^{-3/2}
\label{lr}
\end{equation}
where $\mdot=\dot{M}/\dot{M}_{\rm Edd}$ is the dimensionless accretion
rate ($\dot{M}_{\rm Edd} = L_{\rm Edd}/\epsilon c^2$, $\epsilon=0.1$
is the efficiency of standard accretion, and $L_{\rm Edd}$ is the
Eddington limit), $r_4$ is radius in units of $10^4 R_S$, with $R_S =
2 GM_{\rm BH}/c^2$, the \sch radius for BH of mass $M_{\rm
BH}$. Estimate \ref{lr} was made assuming that the gas is at virial
temperature ($T_{\rm vir}(R) \equiv G\mbh m_p/2k_B R = 2.7\times
10^{12} r^{-1}$ K; $r\equiv R/R_S$). 

According to equation \ref{lr}, if $\lambda > R$ at some radius $R$,
then it is also true for all internal radii unless the accretion rate
$\dot{m}$ increases faster than $\propto r^{-3/2}$ with decreasing
$R$.  Baring that, one can define saturation radius, $R_{\rm sat}$,
via condition $\lambda/R=1$:
\begin{equation}
R_{\rm sat} \simeq 2\times 10^3 R_S
\left[\frac{\alpha}{\mdot}\right]^{2/3}\;.
\label{rsat}
\end{equation}
Below we shall explicitly assume that the heat flux is saturated. The
domain of applicability of our solutions is thus $R \le R_0$, where
\begin{equation}
R_0 \equiv \hbox{min} [R_c, R_{\rm sat}]\;.
\label{r0}
\end{equation}

Formally, our treatment is valid only for non-relativistic
electrons. However, for such high temperatures the electron mean free
path in the cold gas is so large that its exact value does not
particularly matter as long as most of the heat flux is liberated deep
in the cold gas and is re-radiated away, allowing condensation of the
hot gas.

\subsection{Geometry and assumptions}\label{sec:geometry}

The hot flow is sandwiching a very cold and thin disk located in the
midplane (see Figure \ref{fig:geometry}).  As already discussed in \S
\ref{sec:vertical}, a very thin (compared with $R$) transition layer
develops between the corona and the conduction-heated layer of the
disk. For our present purposes it is sufficient to approximate both
the disk and the layer as infinitely thin. Since the cold disk viscous
time is very long, the disk can be treated as stationary. Under these
assumptions the transition layer and the disk enter the dynamical
equations for the hot flow as boundary conditions only.

Due to thermal conduction into the disk, we expect the hot corona to
be cooler than ``normal'' NRAFs that are approximately at the virial
temperatures far from the inner flow region (e.g. BB99; Narayan 2002).
This makes the thermal pressure gradient not important in the radial
momentum equation. The hot flow is thus in a roughly Keplerian
rotation. Similarly, thermally driven winds are weak.

\subsection{Equations for the hot flow}\label{sec:equations}

We now write {\em height-integrated} equations for a hot rotating flow
with a possible mass exchange. Our equations are best understood in
comparison with those of the well known solution obtained by Narayan
\& Yi (1994; hereafter NY94; see their eqs. 1-4), and the equations of
MMH94. We begin with the mass conservation equation. In a one-phase
accretion flow with no winds or mass deposition, the equation is
written as $\partial/\partial R\; [R H \rho v_R]=0$, or $\dot{M}(R)
\equiv 4\pi R H \rho v_R = $~const. Here $\dot{M}(R)$ is the radial
accretion rate, $v_R$ is the radial flow velocity ($v_R>0$ for
accretion), $\rho$ is the height-averaged density in the hot flow. The
vertical scale height $H$ is introduced through the hydrostatic
balance as $H=c_s/\Omega_K$ where $\Omega_K$ is the Keplerian angular
velocity. Taking into account the exchange of mass in the vertical
direction, the mass conservation equation is:
\begin{equation}
\frac{\partial}{\partial R}\;\left[R H \rho v_R\right]= \rho v_z R\;,
\label{massc}
\end{equation}
where $z$ is vertical coordinate with $z=0$ at the cold disk midplane.
Since the radial pressure term is small, the radial momentum equation
(eq. 2 in NY94) is trivially satisfied.  As the cold disk is also
Keplerian, there is no exchange of specific angular momentum between
the two flows and the angular momentum conservation equation (3 in
NY94) is unaltered. With $\Omega=\Omega_K$ we get
\begin{equation}
\dot{M} = 4\pi R H \rho v_R = \frac{12\pi \alpha}{R \Omega_K}
\frac{\partial}{\partial R}\;\left[ \rho c_s^2 R^2 H \right]\;.
\label{vr}
\end{equation}

The entropy equation should include (in addition to the usual terms)
the thermal conduction flux, $\fsat$, and the hydrodynamical flux of
energy in the vertical direction. In the limit of subsonic vertical
flows, $v_z^2\ll c_s^2$, we write
\begin{equation}
Q_+ + \rho v_z \left[ \;\frac{5}{2}c_s^2\; + \;\frac{GM_{\rm
BH}}{2R}\; \left(\frac{H}{R}\right)^2 \right] \; = Q_- + \fsat \;.
\label{eflux}
\end{equation}
On the left hand side of the equation, $Q_+$ is the vertically
integrated viscous heating, and the second term is the mechanical
energy flow in the vertical direction. $Q_-$ is the integrated
radiative cooling inside the corona:
\begin{equation}
Q_- = \Lambda(T_h) \left(\frac{\rho}{\mu}\right)^2 H
\label{qminus}
\end{equation}
where $\mu \simeq m_p$, $T_h$ is the temperature of the hot flow, and
$\Lambda(T_h)$ is the usual optically thin cooling function. The
cooling term due to the thermal conduction is just the saturated
thermal conduction flux, $\fsat$.

Our entropy equation (\ref{eflux}) is the height-integrated version of
that in MMH94 (see their equation 8), with the following
exceptions. As explained in \S \ref{sec:geometry}, we neglect winds
here because we are mostly interested in cooler condensing
solutions. Thus their side-way expansion term, important for $z\simgt
R$ (see their eq. 5), is not included. Hence the last term on the
right hand side of equation 8 in MMH94 is neglected. The radial
entropy flow term (the third line in equation 8 in MMH94) is neglected
for the same reason as winds. For Keplerian rotation, the conversion
of gravitational energy into heat by $\alpha$ viscosity yields the
following heating rate,
\begin{equation}
Q_+ = (9/2)\, \alpha \rho\, c_s^3\;.
\label{qplus}
\end{equation}
It is important to note that both the viscous heating and the thermal
conduction cooling (equation \ref{fsat}) scale linearly with $\rho
c_s^3$. After some simple algebra and using $GM_{\rm BH}H^2/2R^3 =
c_s^2/2$, we find from equation \ref{eflux} that the condensation
velocity is
\begin{equation}
v_z = \left\{{\cal M}_c + \frac{\Lambda(T_h)\rho H}{3 \mu^2
c_s^3}\right\}\, c_s\;,
\label{vz}
\end{equation}
where ${\cal M}_c$ is the Mach number of the condensation velocity in
the important limit of negligible radiative losses:
\begin{equation}
{\cal M}_c = \frac{5}{3}\phi - \frac{3}{2}\alpha\;.
\label{mc}
\end{equation}
Note that had we not neglected radial advective cooling, then writing
it as of $Q_{\rm adv} = f_{\rm adv} Q_+$, where $f_{\rm adv} \le 1$ is
a parameter (NY94), the only necessary change to equation \ref{mc}
would be to change $\alpha\rightarrow\alpha(1-f_{\rm adv})$.

We will be interested in the parameter space in which ${\cal M}_c >
0$, i.e. condensation rather than evaporation. Nevertheless, note that
in the opposite limit equations \ref{vz} and \ref{mc} are not very
accurate because we neglected winds. If $\alpha \simgt \phi$, the
viscous heating dominates the energy balance, the flow is hotter, and
it is unlikely that the winds can be rightfully neglected. The
outflowing wind could simply carry the extra heat away thus
evaporating the hot flow itself (as in the BB99 solution) but not the
cold disk.

\subsection{Boundary conditions}\label{sec:boundary}

At $R=R_0$, we set gas temperature $T_h(R_0)$ to a given ($\le 1$)
fraction of the virial temperature there. The density at $R_0$ is
determined from a given $\dot{M}(R_0)$ value once $v_R(R_0)$ is found.
Since $\fsat\propto p$, the gas pressure, we set $\fsat(z=H) = 0$ in
equation \ref{eflux}. As we neglect the mass outflow from the top of
the corona to infinity, we also write $\rho (z=H) \; v_z(z=H)=0$. The
value of the $v_z(z=0)$, i.e. at the bottom of the corona, is to be
found from equations rather than prescribed.

\section{Approximate analytical solution}\label{sec:analytical}

The first term in equation \ref{vz} is a constant while the last one
is proportional to density of the flow, becoming small at low
accretion rates. We shall thus neglect this term in the non-radiative
limit (see \S \ref{sec:paramspace}). Inserting $v_z = {\cal M}_c c_s$
into equation (\ref{massc}) and also substituting $v_R$ on its value
found from equation (\ref{vr}), we arrive at a second order
differential equation that contains two unknown variables, $\rho$ and
$c_s$:
\begin{equation}
\rho R {\cal M}_c  c_s = 3 \alpha \frac{\partial}{\partial
R}\;\left\{\; \frac{1}{R \Omega_K}\; \frac{\partial}{\partial R}\;
\left[\rho c_s^3 \frac{R^2}{\Omega_K}\right]\right\}\;.
\label{long}
\end{equation}

This equation cannot be solved analytically or numerically in a
general case. By introducing $v_z\ne 0$ we added an extra variable to
the accretion flow equations, and the number of independent equations
is now smaller than the number of unknowns. This situation is well
known in analytical NRAF solutions where one has to introduce three
free functions to obtain a solution (BB99).  In our case there is no
specific angular momentum exchange between the flows, and the heat
flux is specified by $\fsat$, therefore we need to parameterize
``only'' one physical quantity. We have certain well motivated
expectations about the radial temperature dependence. For example, for
a pure non-radiative flow such as a NRAF with no thermal conduction
cooling, $T_h(R)\propto R^{-1}$ for large radii. On the other hand,
thermal conduction tends to smooth out temperature gradients (for
example within supernova remnants), and hence in the other
extreme\footnote{In fact we first studied the problem via numerical
simulations in which we indeed observed this nearly constant
$T_h(R)$}, $T_h(R)\simeq $~const. Thus, we set $c_s = c_{0}
(R_0/R)^{d}$, where $0\le d\le 1/2$ and $c_{0}$ is the sound speed at
$R_0$.

If we define $u\equiv \sqrt{R/R_0}$, then equation (\ref{long}) can be
re-written as
\begin{equation}
\rho u^{3 -2 d} = \frac{3 \alpha}{4{\cal M}_c}\; \frac{R_0 c_0^2
}{G M_{\rm BH}}\; \frac{\partial^2 }{\partial u^2}\;\left[\rho
u^{7-6d}\right] \;.
\label{notlong}
\end{equation}
Finally, defining $\tilde{\rho}\equiv \rho u^{7-6d}$, we obtain
\begin{equation}
\frac{\partial^2 \tilde{\rho}}{\partial u^2} = \frac{1}{l_c^2}\;
\frac{\tilde{\rho}}{u^{4(1-d)}}\;.
\label{simple}
\end{equation}
Here $l_c$ is the dimensionless ``condensation length'':
\begin{equation}
l_c^2 \equiv \frac{3 \alpha}{4{\cal M}_c}\;\frac{ c_{0}^2 R_0}{G
M_{\rm BH}} \; = \; \frac{9 \alpha}{20 \phi - 18 \alpha }\;\frac{
c_{0}^2 R_0}{G M_{\rm BH}}\;.
\label{lc}
\end{equation}
We are mostly interested in the case $\phi \simgt \alpha$ and
therefore we shall only explore the $d=0$ case
below.\footnote{Nevertheless we note that (i) approximate solutions
may be obtained for a general value of $d$, and (ii) for $d=1/2$ there
is a scale-free solution.} In addition, if $\alpha\ll \phi$, the
condensation length is small, i.e. $l_c^2 < \alpha/{\cal M}_c \sim
\alpha/\phi \ll 1$. This circumstance facilitates finding an
approximate analytical solution for equation (\ref{simple}):
\begin{equation}
\tilde{\rho} = \;\hbox{const}\; \exp\left[-\frac{1}{l_c u}\right]\;.
\label{approx1}
\end{equation}
Indeed, $d^2\tilde{\rho}/du^2 = \tilde{\rho}\;(-2/l_cu^3 +
1/l_c^2u^4)\simeq \tilde{\rho}/l_c^2u^4$ due to the fact that $1/l_c u
> 1/l_c \gg 1$ for $u<1$.

Returning to the original variables, we explicitly write down the
approximate solution:
\begin{eqnarray}
\rho(R) = \rho_c \left(\frac{R_0}{R}\right)^{7/2}\;
\exp\left[-\frac{1}{l_c}\left(\sqrt{\frac{R_0}{R}}-1\right)\right]\;,
\label{densr}\\
v_R \;=\; \frac{3 \alpha}{2 l_c}\; \frac{c_0^2}{R_0 \Omega_K(R_0)}
\;=\; \sqrt{\frac{\alpha {\cal M}_c}{3}}\;c_0 \; =\; \hbox{const}\ll
c_0\;,
\label{vrr}\\
\dot{M}(R)\;=\; \dot{M}_0 \frac{R_0}{R}
\exp\left[-\frac{1}{l_c}\left(\sqrt{\frac{R_0}{R}}-1\right)\right]\;.
\label{mdotr}
\end{eqnarray}
Here $\rho_c$ is the gas density at $R_0$, and $\dot{M}_0$ is the
accretion rate at that point. Note that $l_c R_0$ is roughly the
radial distance over which the hot flow will condense. At $R_0-R\ll
l_c R_0$, there is little left of the hot flow.

Note also that equation (\ref{vrr}) shows that the radial velocity is
constant and is substantially smaller than the sound speed since we
study cases when both $\alpha$ and ${\cal M}_c\ll 1$. Further, for
$l_c< 1/2$ (recall that we assumed $l_c\ll 1$), the accretion rate
increases with $R$ for all $R< R_0$, as it should for a condensing
flow. Also note that the radial flow velocity is small compared with
the sound speed and yet it is larger than expected for the usual
one-phase hot flows for which $v_{R, \rm visc} \sim \alpha c_s (H/R)^2
\simlt \alpha c_s$.

\begin{figure}
\centerline{\psfig{file=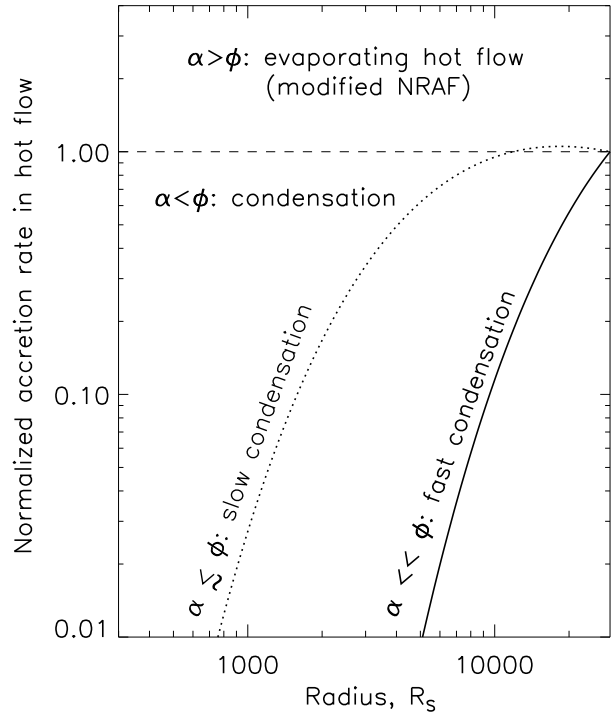,width=.45\textwidth,angle=0}}
\caption{Radial accretion rate profiles in the hot flow normalized on
$\dot{M}_0=\dot{M}(R_0)$ (for $R_0=3\times 10^4 R_S$). The saturated
heat flux coefficient is $\phi=0.2$ for all the cases. The
$\dot{M}/\dot{M}_0=1$ (dashed) curve shows the case when the hot flow
neither evaporates not condenses, which occurs when $\alpha\simeq
\phi$. For $\alpha$ smaller than $\phi$, the hot flow is
condensing. When $\alpha$ is only slightly lower than $\phi$, $l_c\sim
1$, and the condensation is slow (cf. the dotted curve). For $\alpha
\ll \phi$, condensation length $l_c\ll 1$, and the hot flow very
quickly settles down on the cold disk. This latter case is illustrated
by the solid curve for which $\alpha=0.02$. Finally, for $\alpha\simgt
\phi$, the hot flow (but not necessarily the cold disk!) evaporates,
outflowing to infinity as in the BB99 solutions.}
\label{fig:radial}
\end{figure}

Figure \ref{fig:radial} shows $\dot{M}/\dot{M}_0$ as given by equation
\ref{mdotr} for $\phi=0.2$ and two different values of $\alpha$,
i.e. $\alpha = 0.02$ and 0.1 (solid and dotted curves,
respectively). The accretion rates are normalized on their values at
$R=R_0$. The horizontal dashed curve shows the case of no mass
exchange. Note that for the case of very small $\alpha$, $\mdot$
essentially plunges with radius, whereas for $\alpha = 0.1$ (which is
only a factor of 2 smaller than $\phi$), $\mdot\simeq $~const for a
range or radii before it reaches the exponential decline.

\del{The approximate solution was obtained under the assumption that
$\alpha \ll \phi\ll 1$. This guarantees that the condensation velocity
is sub-sonic (i.e. ${\cal M}_c\ll 1$, see equation \ref{mc}). However
qualitative features of the solution should remain valid as long as
$\alpha < \phi$. }

\subsection{Spectra from condensing flows}\label{sec:spectra}

For low luminosity systems, Comptonization in the corona is not
important.  Therefore we expect the coronal spectra to be dominated by
the free-free emission as well as X-ray line emission (at the coronal
ionization equilibrium) for $T_h\simlt 10^8$ K. The single temperature
free-free spectrum is thus the hardest one that can be expected here;
the photon spectral indices are then $\Gamma\simgt 1.4$ or so. The
spectra are softer for lower $T_h$ and also in the case when there is
a broad distribution of temperatures in the corona.

The maximum transition layer temperature is $T_c\simlt 10^5$ K. Gas at
this temperature is a very efficient emitter of UV lines, e.g. that of
He$^+$, Oxygen and other elements. The column depth and thus the
emission measure of the gas at $T\simeq 10^5$ K depends on the exact
value of the non-local thermal conduction heating, $H_{\rm tot}$
(equation \ref{htot}), which depends on the hot flow temperature
mainly.  If the temperature is below $\sim 10^8$ K, then the $\sim
10^5$ K gas will re-radiate a significant fraction of the condensing
flow energy. In this case there should be a strong UV bump in the SED
of the source, along with another bump of comparable magnitude at the
effective temperature of the disk, which is in the infra-red to the
optical frequency range depending on the parameters (black hole mass,
condensation radius, condensation rate).

If the hot flow temperature is greater than $10^8$ K or so at the
point where most of the condensation takes place, on the other hand,
then the situation is qualitatively different. Since $H_{\rm tot}$ is
lower, the layer with $T\sim 10^5$ K exists only due to local thermal
conduction and is much thinner than in the opposite case (NS04). Bulk
of the conduction-heated zone is at temperature $T \simlt 10^4$ K.
Strong Hydrogen Ly$\alpha$, H$\alpha$, H$\beta$, etc., line radiation
is expected from these layers rather than UV emission. Thus in this
case the UV region in the SED of the condensing flow is weak.

As shown in Figure \ref{fig:temp-sketch}, below the conduction-heated
layer there will be also a photo-ionized layer, an analog of an HII
region. This deeper layer is expected to be a strong emitter of
optical lines, especially H$\alpha$, H$\beta$. Therefore note that in
the present model the vertical structure of the transition layer
contains both collisionally and photo-ionized emission regions whose
relative contribution to the optical spectrum will vary depending on
conditions.

Finally, below the optically-emitting zone there will be the optically
thick cold disk. Its effective temperature will be controlled by both
the external energy deposition (the condensing flow plus starlight,
etc.)  and the internal energy dissipation. For LLAGN, this typically
yields $T_{\rm eff} \sim 10^3$ K.

\section{Parameter space}\label{sec:paramspace}

\subsection{Saturated accretion rate}\label{sec:sar}

\begin{figure}
\centerline{\psfig{file=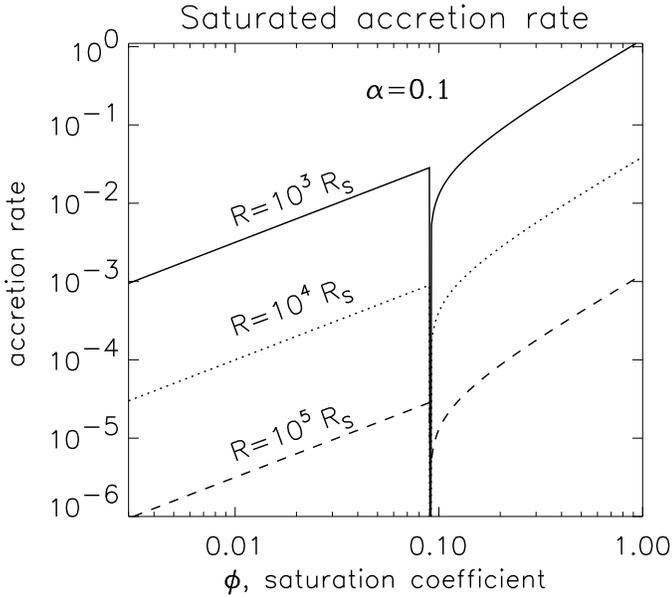,width=.5\textwidth,angle=0}}
\caption{The accretion rate, $\msat$, below which the heat flux in the
hot flow is saturated (collisionless), as a function of the saturation
parameter $\phi$. All the three curves are for $\alpha=0.1$ but
varying radii as labelled above the respective curves. The $\msat$
curve is rather approximate in the region $\phi\simeq \alpha$ where
the two different limits meet (see equation \ref{msat}). Note that for
the largest radius, the hot flows are saturated only for extremely
small values of $\mdot$. On the contrary, for the smallest radius, the
hot flows are in the collisionless regime even for relatively large
$\mdot$ unless $\phi\ll 1$.}
\label{fig:msat}
\end{figure}

First we specify the conditions under which the heat carrying
electrons in the hot flow are collisionless, i.e. the heat flux is
saturated. The estimate is done at radius $R_0$, where we assume that
the hot flow temperature is virial (in \S \ref{sec:global} we discuss
$R < R_0$). Note that for a $\phi < 1$ one has to reduce the electron
mean free path $\lambda$ (equation \ref{lcm77}) to $\lambda'\equiv
\phi \lambda$. Indeed, the saturation coefficient $\phi$ is an ad hoc
attempt (see Appendix A) to account for the reduction of the electron
heat flux due to magnetic fields, plasma instabilities, etc., and to
be consistent one should also reduce the electron effective mean free
path. The critical gas density can then be related to the accretion
rate given that we know the radial velocity of the flow, $v_R$. For
$\phi \simgt \alpha$ we use equation \ref{vrr}, whereas for the case
of $\phi \ll \alpha$ we can simply neglect the thermal conductivity
altogether and use the usual viscous radial velocity scaling (see the
first paragraph in \S \ref{sec:rad}). Thus we arrive at the following
condition for the thermal conductivity of the hot flow to be
collisionless:
\begin{equation}
\mdot < \msat = \cases{ 0.1 \; \sqrt{\alpha {\cal M}_c}\; \phi\;
r_4^{-3/2}, & if $\phi\simgt \alpha$\cr 0.1 \; \alpha \; \phi\; r_4^{-3/2},
& if $\phi\simlt \alpha$.\cr}
\label{msat}
\end{equation}
Figure \ref{fig:msat} shows $\msat$ for three different radii and a
same value for $\alpha$-viscosity parameter, $\alpha=0.1$. The sharp
drop in the curves around $\phi \simeq \alpha$ is the artifact of our
assumption $\phi\gg\alpha$ that we used to build an approximate
analytical solution in \S \ref{sec:analytical}; in the reality a
numerical integration of the equations would yield a smoother
transition between the two cases given in equation \ref{msat}.  For
radii as large as $10^5 R_S$, the hot flow must be accreting at a
dimensionless rate smaller than $\sim 10^{-3} \phi$ for the heat flux
to be saturated. However the flow is almost always saturated for
``small'' radii such as $10^3 R_S$.

\subsection{Radiative condensation}

As noted in \S \ref{sec:clascond}, spontaneous condensation may be
taking place even in an originally one-phase system without any
thermal conduction if the in-situ radiative cooling is too large. Hot
accretion flows suffer a catastrophic collapse when the cooling
exceeds the viscous heating. For a NRAF, this occurs at accretion
rates exceeding $\mdot_{\rm crit}\sim \alpha^2 r_4^{-1/2}$
(e.g. Narayan \& Yi 1995; although exact results depend on the cooling
processes and geometry of the flow: Esin 1997). In the presence of the
cold accretion disk below the hot flow, the additional thermal
conductive cooling causes the latter to condense at a smaller
pressure. In \S \ref{sec:clascond} we showed that this shifts the
value of the critical radiative condensation rate $\mrad$ downwards
by a factor of order unity only, i.e. $\mrad\sim \dot{m}_{\rm crit}$
(but see \S \ref{sec:global} below).

\begin{figure*}
\begin{minipage}[b]{.47\textwidth}
\centerline{\psfig{file=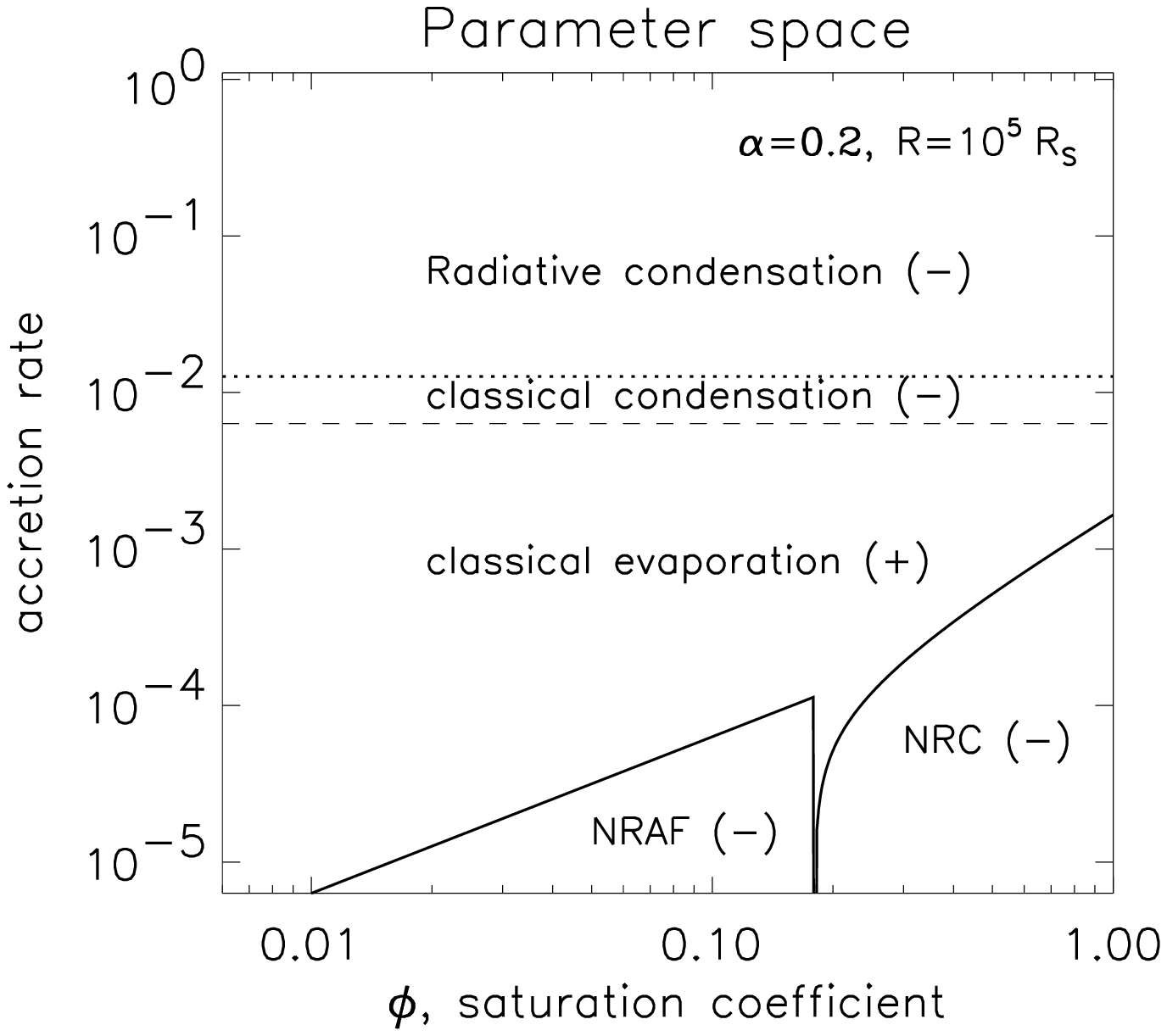,width=\textwidth,angle=0}}
\end{minipage}
\begin{minipage}[b]{.47\textwidth}
\centerline{\psfig{file=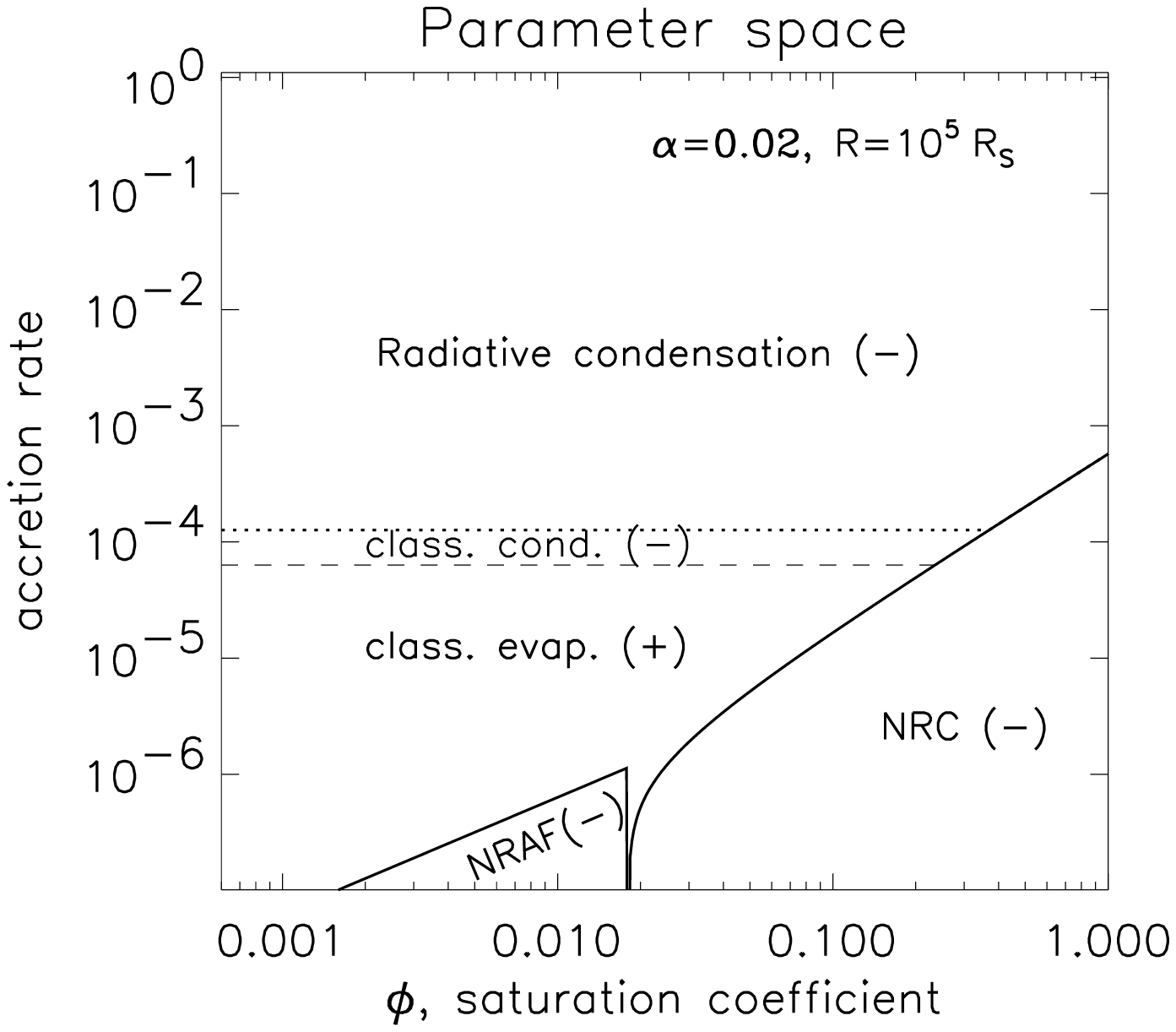,width=\textwidth,angle=0}}
\end{minipage}
\caption{Left panel: The types of solutions occurring in the two-phase
accretion flows at radius $R=10^5 R_S$ and for a ``large'' value of
$\alpha$-parameter ($\alpha=0.2$), as a function of the dimensionless
accretion rate and the saturation parameter $\phi$. The signs $+$ or
$-$ reflects the sign of the mass exchange between the disk and the
hot flow. See text in \S \ref{sec:division} for a detailed discussion
of the Figure. Right panel: same as the Left one but for $\alpha =
0.02$.}
\label{fig:a0.2r10}
\end{figure*}

Now, if $\mdot < \msat$, i.e. the heat flux is collisionless, and the
thermal conductivity is large ($\phi \simgt \alpha$), we use the
solutions obtained in \S \ref{sec:analytical}. The radiative cooling
contribution to the condensation can be neglected when $\Lambda(T)
\rho H /(3 \mu^2 c_s^3) \ll {\cal M}_c$ (see equation \ref{mc}).
Evaluating this relation at $R=R_0$, assuming the free-free cooling,
$\Lambda(T)= 2.4\times 10^{-27} T^{1/2}$, one finds the accretion
rate, $\mrad$, at which radiative losses start to dominate the energy
balance in the hot condensing flow:
\begin{equation}
\mrad = 6 \alpha^{1/2} {\cal M}_c^{3/2} r_4^{-1/2}\;.
\label{mrad}
\end{equation}
If $\mdot < \msat$ but $\phi \ll \alpha$, then the thermal conduction
can be neglected and we again estimate that $\mrad \sim \alpha^2
r_4^{-1/2}$, as in a pure NRAF case. Combining together these results,
we obtain the accretion rate above which condensation is driven by the
in-situ radiative losses in the hot flow:
\begin{equation}
\mrad = \cases{6 \alpha^{1/2} {\cal M}_c^{3/2} r_4^{-1/2}, &if
$\phi\simgt \alpha\;$ and $\;\fcl > \fsat$\cr 
\alpha^2 r_4^{-1/2}, &if $\phi\simlt\alpha\;$ or $\;\fcl < \fsat$\cr}
\label{mrad}
\end{equation}

\subsection{Division of the parameter space}\label{sec:division}

We can now divide the parameter space into different regions by the
type of the flow that occurs there. The division is governed by the
following principles:

\begin{itemize}

\item $\mdot < \msat$, $\mdot < \mrad$, $\phi \simgt \alpha$:
Non-Radiative Condensation (NRC for short), the solutions obtained in
this paper.

\item $\mdot < \msat$, $\mdot < \mrad$, $\phi \simlt \alpha$:
Non-Radiative Accretion Flow (NRAF), with a weak (since $\phi$ is
small) collisionless condensation. The flow structure is similar to
the usual NRAFs except the flow should be somewhat cooler.

\item $\mdot > \msat$, $\mdot < \mrad$ Classical evaporation of the
disk by the hot flow (e.g. MMH94). ``Classical'' reflects the fact
that the heat flux is in the classical, collision-dominated, regime.

\item $\mdot > \mrad$: Independently of other parameters, the hot
corona is condensing onto the cold disk because of a too strong in-situ
radiative cooling.

\end{itemize}

To illustrate the resulting division in the $\mdot-\phi$ parameter
space, first consider large radii, such as $R=10^5 R_S$, and large
$\alpha$ (Figure \ref{fig:a0.2r10}, Left panel). For each of the
segments in the parameter space, the respective type of solution is
indicated, and the signs ``+'' or ``$-$'' indicate the sign of the
mass exchange between the cold disk and the hot flow; ``+'' stands for
the evaporation of the cold disk.

At the largest accretion rates (above $\mdot\simeq 0.01$), the flow
condenses due to the radiative losses in the hot flow itself. Slightly
below that there is a narrow zone of the classical condensation, which
is enabled by the radiative losses inside the hot flow and the
transition region.

At yet lower accretion rates, the radiative cooling in the hot flow
and the transition layer becomes too weak. The excess energy is used
to evaporate the cold disk, as first discussed by MMH94. The hot flow
also drives a thermal wind to infinity (MMH94), and in many respects
is similar to the NRAF solutions (e.g. BB99; Narayan 2002), although
the two-temperature assumption typical for NRAF is not required for
these evaporating flows.

Next, below $\mdot=\msat$, lie the parameter space where the heat flux
is collisionless. As we found in this paper, as long as $T_h\simgt
T_{\rm cond}$, the heating of the transition layer by the hot
electrons can be counter-balanced by the local radiative losses, and
thus the hot flow is condensing. When $\phi \simgt \alpha$, the
approximate solutions developed in \S \ref{sec:analytical} apply. The
hot flow is condensing due to radiative cooling in the transition
layer. Unlike the classical condensation regime, however, the
transition layer is very much cooler in this case ($T\simlt 10^5$ K;
see \S \ref{sec:vertical}), and the hot electrons are in a
free-streaming collisionless regime. The hot flow can be completely
quenched (terminated) in this regime as $\dot{M}(R \rightarrow 0)
\rightarrow 0$.

Finally, at $\phi \simlt \alpha$ and $\mdot<\msat$, we have an analog
of a hot non-radiative corona that we simply call a NRAF (although
once again, it does not have to be a two-temperature flow as
such). While the hot flow is still condensing in this case, the
process is slow since $\phi < \alpha$, and thus the hot flow will not
be terminated as in the NRC case. Therefore, the NRAF (--) case may be
quite analogous to the usual NRAFs except for the additional (weak)
cooling and condensation provided by the thermal conduction.

\begin{figure*}
\begin{minipage}[b]{.47\textwidth}
\centerline{\psfig{file=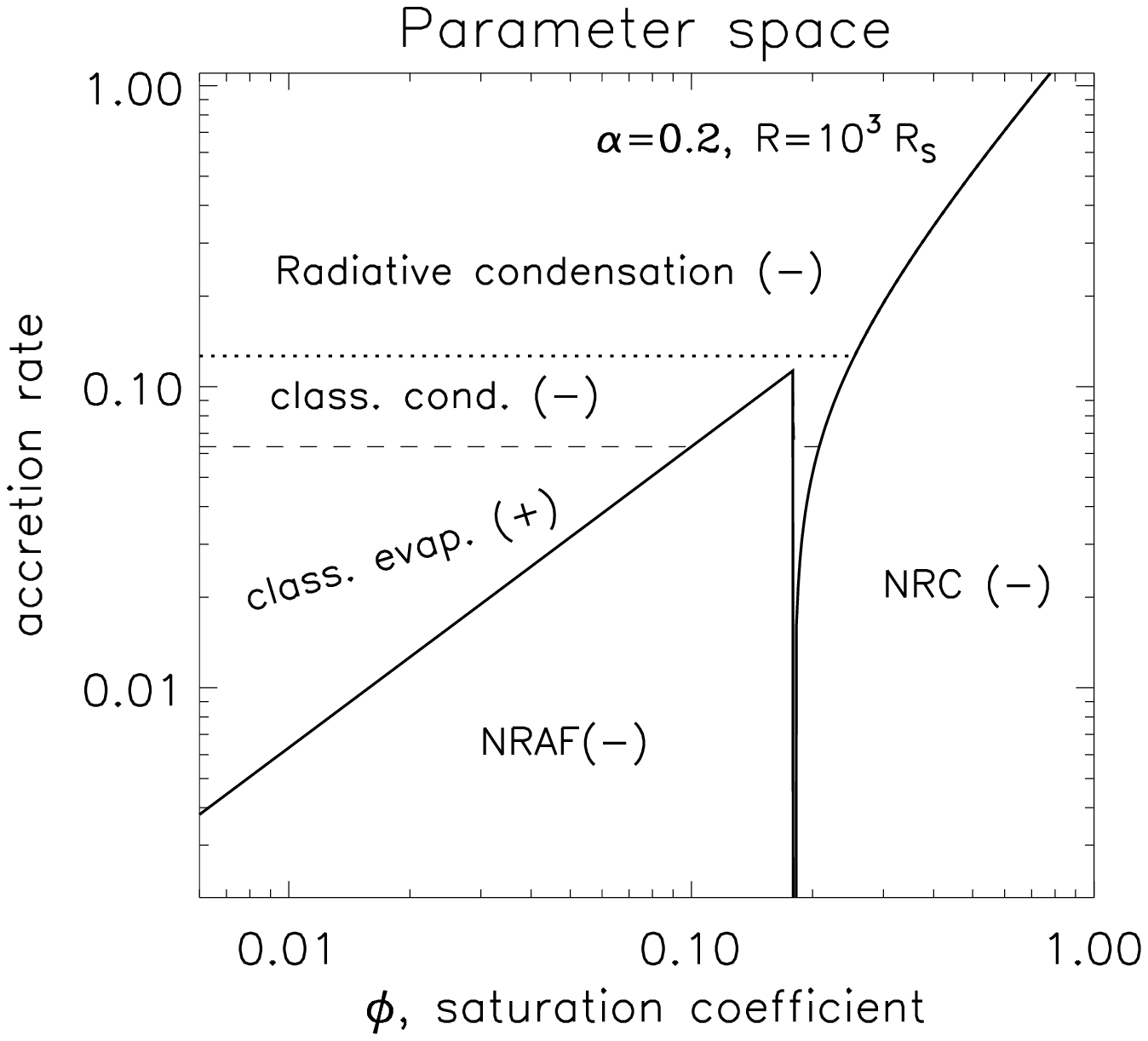,width=\textwidth,angle=0}}
\end{minipage}\hfill
\begin{minipage}[b]{.48\textwidth}
\centerline{\psfig{file=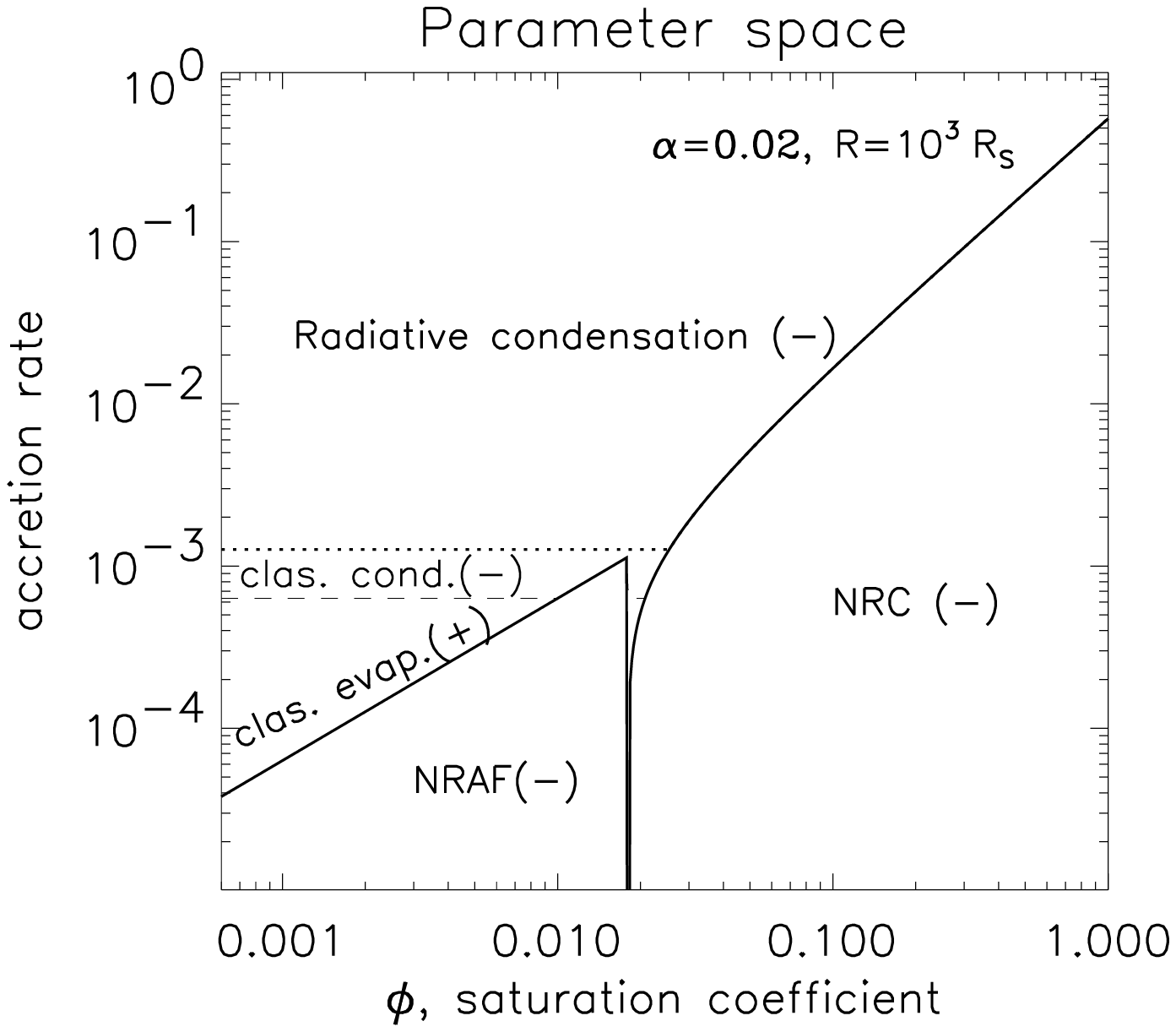,width=\textwidth,angle=0}}
\end{minipage}
\caption{Same as Figure \ref{fig:a0.2r10}, but for $R=10^3 R_S$.  Note
that because of the much higher virial temperature at $R=10^3 R_S$,
the thermal conductivity in the hot flow is collisionless for much
larger accretion rates than in Figure \ref{fig:a0.2r10}. Thus the
region of applicability of the NRC solutions grows, while the
``classical evaporation'' zone shrinks.}
\label{fig:r0.1}
\end{figure*}

Consider now smaller radii ( e.g., $R=10^3 R_S$ in Figure
\ref{fig:r0.1}). The hot flow is now much hotter, and thus the column
depth penetrated by hot electrons is much larger. Therefore the flow
is much more likely to be in the collisionless regime. For
sufficiently small radii, the hot flow can be both collisionless and
radiative along the line separating the radiative condensation and the
NRC solutions in the upper right corner of both panels in Figure
\ref{fig:r0.1}. This means that for such small radii the classical
evaporation regime completely disappears for large values of
$\phi$. As can be seen in Figure \ref{fig:r0.1}, for $\phi\simgt
\alpha$, the hot flow condenses for any $\mdot$, either due to in-situ
radiative losses (high $\mdot$) or due to radiative losses in the
transition layer (low $\mdot$) as discussed in \S\S\,
\ref{sec:vertical} \& \ref{sec:analytical}.

\subsection{On global nature of realistic two-phase
flows}\label{sec:global} 

As stated in the beginning of \S \ref{sec:paramspace}, so far we have
only considered the parameter space for condensation or evaporation at
the outer radius of the hot flow, $R=R_0$, where we assumed
$T_h=T_{\rm vir}$. However at smaller radii, where the radial
(in-flow) velocity significantly decreases compared to the free-fall
value, one expects that the thermal conduction will be much more
efficient in cooling the hot flow to $T_h \ll T_{\rm vir}$.

Let us determine how the value of $\mrad$ changes in this case. The
integrated heating rate, $Q_+$, is the rate at which the inflowing gas
liberates its gravitational energy, and hence it only depends on the
accretion rate, $\mdot$, and is independent of the coronal
temperature. The radiative cooling rate, however, scales as $Q_- \sim
\Lambda(T_h) \rho^2 H$. Since $H/R \simeq \sqrt{T_h/T_{\rm vir}}$ for
accretion flows, and $v_R \simeq \alpha \Omega_K R (H/R)^2$, we have
$\rho \propto \mdot (T_{\rm vir}/T_h)^{3/2}$. Taking $\Lambda(T_h)
\sim$~const, we see that the radiative cooling rate approximately
equals the heating rate at
\begin{equation}
\mrad \sim \alpha^2 r_4^{-1/2} \left( \frac{T_h}{T_{\rm
vir}}\right)^{5/2}\;,
\label{mradmod}
\end{equation}
which may be significantly smaller than that given by equation
\ref{mrad}. Evidently, the influence of the thermal conduction on the
collapse of a hot flow is very significant in this example in the
cumulative sense. While a hot one-phase accretion flow with $T_h\simeq
T_{\rm vir}$ may be thermally stable to collapse (spontaneous
condensations), the two-phase hot flow with same accretion rate would
be much cooler, $T_h\ll T_{\rm vir}$, much denser, and would thus
collapse onto the cold disk. The detailed calculations by Liu et
al. (2004) show that this is exactly the case. These authors find a
condensing solution for $\alpha=0.1$ and accretion rate $\mdot\sim
3\times 10^{-3}$, which would be somewhat too low for the classical
condensation. However their vertical temperature profiles show that
the coronal temperature is only $T_h\sim 0.3 T_{\rm vir}$, making
$\mdot < \mrad$ according to equation \ref{mradmod}, and enabling
condensation.

Even more complex situation may exist when the gas is evaporating at
large radii but condenses at small radii, or the flow is in the
classical thermal conduction limit at large radii and is yet in the
saturated conduction limit at small $R$. This complexity of the
two-phase flows with thermal conduction is not something new, however,
as accretion flow equations are well known to allow more than one
self-consistent solution at same $\dot{M}$ and radius (e.g. Abramowicz
et al. 1995).

\section{Appearance of condensing flows}\label{sec:obs}

From the disk ionization instability theory (for reviews see Cannizzo
1998, Lasota 2001), it is well known that ``cold'' accretion disks may
be in two states: (i) active, that is accreting, in which case the
disk is approximately in a steady state, with the accretion rate equal
to a constant independent of radius. (ii) quiescent or inactive, when
most of the hydrogen is neutral. The disk then accumulates the mass
and the accretion rate strongly decreases with decreasing radius $R$.

In the case of an active disk, we expect that a steady-state will be
reached in which the mass deposited by the condensing flow is accreted
with time onto the black hole. The luminosity of the inner accretion
flow is then much larger than that of the condensing flow at large
radii. Accordingly, the latter is a minor detail in the overall
spectrum of the source. This situation is more likely to realize for
the classical thermal condensation that takes place at a relatively
high accretion rate than the non-radiative condensation.

In the opposite case of a quiescent disk, the situation is completely
different since then the luminosity of the system is much smaller and
the condensing flow may be the primary radiation source. This
astrophysically interesting case is more likely for the non-radiative
condensation, and is discussed below.

\subsection{Underluminous accretion flows}\label{sec:inactive} 

If the cold disk is inactive, i.e. barely accreting, in analogy to
quiescent disks in binary transient systems (e.g. see Cannizzo 1998;
Lasota 2001), then the mass condensing on it may be stored there for
very long times. Indeed, the viscous time scale at radius $R$ is
\begin{equation}
\tvisc \sim \; 3 \times 10^8 \; \hbox{years}\quad \frac{M_8
r_4^{1/2}}{\alpha_2 T_3}\;,
\label{tvisc}
\end{equation}
where $M_8$ is the black hole mass in units of $10^8$ years,
$\alpha_2$ and $T_3$ are the cold disk viscosity and the temperature
in units of $0.01$ and $10^3$ Kelvin, respectively.  For example, for
\sgra, with $M_8\sim 0.03$ and with $T_3\sim 0.1$ (e.g. Nayakshin et
al. 2004), this corresponds to $10^8$ years, far longer than the age
of the bright young stars estimated at $\simlt 10^7$ years (e.g. Ghez
et al. 2003). For a black hole mass of $\sim 10 \msun$, $\tvisc \sim
30$ years, a time relatively long for binary systems.

For inactive disks, the accretion rate through the inner boundary,
$\dot{M}_i$, is much smaller than that at large radii (e.g. Nayakshin
\& Svensson 2001), $\dot{M}_0$.  Some rough estimates are
possible. Suppose the condensing gas is deposited onto a ring with
radius $R=R_0$, and that the initial mass of the cold disk,
$M_d(t=0)$, is negligible small. Thus the disk mass at time $t$ is
$M_d(t)\simeq t \dot{M}_0$. The time needed for the matter to diffuse
to the innermost stable orbit is exactly the viscous time. The rate at
which the black hole consumes the disk can be estimated as (see, e.g.,
\S 5 of Frank et al. 2002)\footnote{Exact results will depend on the
run of the cold disk temperature with radius, but the dominant
exponential factor will be always similar}
\begin{equation}
\dot{M}_i(t) \sim \frac{M_d(t)}{\tvisc}
\exp\left[-\frac{\tvisc}{t}\right]\;\sim\; \dot{M}_0\; \frac{t}{\tvisc}
\exp\left[-\frac{\tvisc}{t}\right]\;.
\label{min}
\end{equation}
This equation shows that as long as $t \ll \tvisc$, the innermost
accretion rate is very small compared with that of the hot condensing
flow. The expected luminosity of the innermost disk is $L_i \sim 0.1
\dot{M}_i c^2$. Comparing this with the luminosity of the condensing
flow at $R=R_0$, $L_{\rm cond} \sim (3 R_S/R_0) 0.1 c^2\; \dot{M}_0$,
\begin{equation}
\frac{L_i(t)}{L_{\rm cond}} \; \sim \;\frac{\dot{M}_i(t)}{\dot{M}_0}\;
\frac{R_0}{3 R_S}\; \sim \; \frac{R_0}{3 R_S} \; \frac{t}{\tvisc}\;
\exp\left[-\frac{\tvisc}{t}\right]\;.
\label{lcor}
\end{equation}
Even if $R_0 \gg 3 R_S$, the luminosity of the inner disk is small as
long as $t\ll \tvisc$.

Therefore, we see that the hot condensing flow in coupe with an
inactive cold disk are ``underluminous'' when compared with accretion
flows that carry the gas directly into the black hole (i.e. assuming
the standard 10\% efficiency of the mass to radiation conversion for
accretion). The effect is of course a time dependent one; upon
averaging over $t\gg \tvisc$, the average luminosity of the two-phase
flow equals that of the standard accretion flow (Shakura \& Sunyaev
1973).

\subsection{Missing or truncated inner disks}\label{sec:missing}

If the cold disk is truly inactive, only the area of the disk actually
covered by the hot flow will be emitting. If one estimates the disk
extent via the velocity width of H$\alpha$ line or via the SED of the
source, then it may appear that there is no disk at small radii
because most of the hot flow does not reach there. While it is quite
possible that the inner disk is missing altogether, it may also be in
the inactive state where it emits very little to be detectable, as
explained above.

\subsection{SED for non-radiative condensation}\label{sec:sed}

In the two-phase flows studied here, most of the thermal energy
resides in the hot corona. However, in the non-radiative case, the
corona is cooled by thermal conduction instead of radiation. Owing to
this, the radiation output of the condensing flow is dominated by the
radiation of the underlying disk.  X-ray emission for such flows
should always be week compared with the thermal-like bump in the
infrared to UV frequency region (depending on the black hole mass and
other parameters).

\subsection{SED for radiative condensation}\label{sec:sedrad}

If the hot gas condensation is initiated by the radiative losses in
the corona itself (\S \ref{sec:clas}), then most of the hot flow
energy is emitted in X-ray frequencies. Baring the case of a very
optically thin cold accretion disk, about half of this radiation is
absorbed in the disk and is re-emitted in the infra-red and/or optical
frequencies. Thus in this case the SED of the source should contain a
similar amount of power in the X-ray and the disk components.

Clearly, a case intermediate between this and that considered in \S
\ref{sec:sed} exists, when the hot flow persists for a decade or more
in radius before succumbing to the condensation process (see end of \S
\ref{sec:global}).

\subsection{Broad double-peaked emission lines}\label{sec:halpha}

As we discussed in \S \ref{sec:spectra}, the gas sandwiched between
the corona and the cold disk is a strong emitter of optical and UV (if
hot enough) lines. X-rays from the corona also contribute to radiative
ionization and emission of optical/UV lines.  Also note that the hot
completely ionized coronal gas, settling down on the disk, will have
to recombine, again producing emission lines. Therefore a generic
prediction of our model is the presence of strong emission lines that
should be shifted and broadened due to disk rotation, resulting in
double-peaked line profiles. We expect the lines to trace a broad
range of conditions and hence be a mix of photo-ionized and
collisionally ionized cases.

\subsection{Fluorescent Fe K$\alpha$ lines}\label{sec:iron}

Here we have considered the hot X-ray emitting flow at large ($R\sim
10^3 - 10^5 R_S$) distances from the black hole. If such a flow
condenses and liberates most of its energy at these large distances,
when clearly any Fe K$\alpha$ line due to reprocessing of the
radiation in the cold disk would be ``narrow'', in accord with the
observations (e.g. Ptak et al. 2004; Dewangan et
al. 2004). Furthermore, because the inactive disks may be powered by
condensation of the hot gas rather than by the internal disk
accretion, the column density of these may be much lower than that
deduced from the cold disk equations with the accretion rate tuned to
match the optical/UV output of the source. Thus the inactive disks
studied here may actually be Thomson-thin, producing no detectable Fe
K$\alpha$ line at all.

\section{Conclusions}\label{sec:conclusions}

Here we studied the physics by which a hot flow above a cold accretion
disk could condense. Such a condensation process is of a large
practical importance for accretion in AGN and X-ray binaries. Without
means to cool, the hot gas is an extremely inefficient kind of fuel
for the black hole (e.g. BB99). Not only the gas radiates little, it
also {\em accretes} little (for recent numerical simulations see Proga
\& Begelman 2003). In contrast, if the hot gas condenses onto a cold
accretion disk, the gas may temporarily loose its viscosity, i.e. the
ability to accrete efficiently, but it remains tightly bound to the
black hole. With time, when accretion through the cold disk is
restarted, the hot gas will accrete onto the black hole. Thus our
condensing scenario for the accretion of the hot gas onto the black
hole is much more efficient in feeding the hot gas into the black hole
than the non-radiative hot flows (BB99, Narayan 2002).

We have found two distinct condensation regimes: (i) radiative or
classical, which takes place when the radiative cooling term in the
corona is comparable with the viscous heating term (see \S
\ref{sec:clas}) and the thermal conductivity is classical; and (ii)
non-radiative. When the flow density is very low and the radiative
losses are negligible. However the hot particles have very long mean
free paths that allow them to penetrate the cold gas directly. Their
energy is then re-radiated by the dense cold layers (see \S
\ref{sec:vertical}). The cold disk thus serves as a cooling plate,
radiator, for the hot flow. This second condensation regime is not
expected based on the previous literature that employed the classical
or the saturated heat flux formulations.

We also presented a simple analytical solution describing the radial
structure of the hot flow for the non-radiative condensation. The
solution is obtained under the assumption of hot flow temperature
constant with radius and is meant to be an example only. The detailed
structure of the flow is not as important for the flow energetics as
long as one can clearly define a radius $R'$ where most of the corona
condenses. As the hot gas condenses, the gravitational energy released
per unit mass of hot gas is $\sim G\mbh/2R'$. If the cold disk is
inactive, then most of the condensed mass is stored in the disk until
the disk becomes massive enough (e.g. Siemiginowska, Czerny, \&
Kostyunin 1996; but note that galaxy mergers, etc., may be another
``direct'' way of triggering accretion outbursts in the inner parsec
from the BH).  Thus the luminosity of the condensing flow (while the
disk is inactive) is quite small compared with that expected for
standard accretion flows (\S \ref{sec:inactive}).

Nuclei of LLAGN is an example where the two-flow geometry is natural
-- e.g. see Fig. 2 of Ho (2003) and note that in reality the hot gas
probably fills the whole available space because it is too hot to be
confined (BB99). Currently, the bump-like infra-red feature seen in
the SED of LLAGN is thought to be due to the {\em accretion} through
the disk. Assuming that the flow in the disk is continuous down to the
last stable orbit however leads to a paradox since then one would
expect a far brighter source. In contrast, if the cold disk is powered
by the condensing corona and there is no accretion in the disk itself,
there is then no problem with the source being too dim.  The current
``non-radiative'' condensation mechanism is thus somewhat
inconspicuous in that the radiative output of such a flow is dominated
by the emission of a warm atomic and ionized gas and not by the energy
source -- the hot gas.

A similar picture may be relevant for quiescent disks in transient
binary systems if not all of the accretion stream radiatively
condenses in the hot spot (the hot spot emission is quite weak in many
systems; see references in Lasota 2001, his \S 7). Thus in some
systems the ``disk'' continuum and line emission may be excited by
shocks in the hot spot, whereas in others (with a weaker hot spot),
the emission may be produced by the diffuse hot flow condensing onto
the disk.

Since the accretion flow may span a broad range of radii, it is
possible that different types of mass exchange solutions realize at
different radii. For example, at a high enough accretion rate, the hot
flow may be condensing radiatively at large radii, whereas at small
radii, where the accretion rate is significantly reduced, it may
condense by the non-radiative mechanism. It is equally possible for
the accretion flow to be evaporating at large radii and condensing at
small ones. Therefore we should keep in mind that in general a mix of
the solutions is possible. This ``unfortunate'' multi-phase complexity
of accretion flows is not something unique to the accretion process;
after all the interstellar medium has a multi-phase structure too.

Finally note that due to processes not taken into account in the
standard disk instability model (e.g. Cannizzo 1998), for example
photo-ionization of the disk upper layers by starlight and X-rays;
stellar impacts (Nayakshin et al. 2004), etc., there will always be a
finite (weak) accretion onto the BH. This weaker accretion may power
the jet emission.

Our results also indicate that a left alone cold disk will {\em not}
necessarily have to evaporate into a hot corona at a low $\mdot$ via
the classical thermal conductivity (MMH94). Indeed, when a hot corona
just starts developing above the disk, the coronal accretion rate
initially is very low and it is thus in the collisional, not
classical, regime.

\section{ACKNOWLEDGMENTS}

We thank Rashid Sunyaev for his careful reading of an earlier draft of
the paper, and the many useful suggestions and comments that he made
that substantially improved this work.  We also acknowledge very
useful discussions with F. Meyer, E. Meyer-Hofmeister, B. Liu,
H. Spruit \& K. Dullemond. Finally, C. McKee is thanked for several
illuminating discussions during the author's visit to UC-Berkeley.

\del{that were
instrumental in advancing his understanding of the physics of thermal
conductivity.}

\appendix

\section{Notes on complications not considered here}

\paragraph*{Magnetic fields} 
Magnetic fields are not explicitly introduced in this paper. The
reason is pragmatic -- magnetic fields in accretion flows are known to
be turbulent and time-dependent (e.g. Balbus \& Hawley 1991) and there
seems to be no hope of obtaining an analytical solution such as that
found in \S \ref{sec:analytical} when thermal conduction and fields
are both considered. The presence of even a very weak (far
sub-equipartition) magnetic field strongly reduces the thermal
conductivity in direction perpendicular to the field lines. The usual
approach is to set the conductivity to zero in that direction (e.g.,
see Balbus 1986). Thus, the effective value of $\phi$ will be reduced
by any non-zero magnetic field, but an essential point is that {\em
this reduction is a function of the field topology but not the field
strength.} The main unknowns in this regard are the field correlation
length and the angular distribution of the field directions near the
interface between the hot and the cold flows.

The value of the effective viscosity coefficient $\alpha$ is also a
function of the magnetic field, but {\em a different function}. While
the correlation length and the angular directions are again important,
the absolute magnitude of $\alpha$ is nearly linearly proportional to
the magnetic field pressure (e.g. Balbus \& Hawley 1991). Furthermore,
the effective value of $\alpha$ should be obtained by averaging over
the whole hot flow, whereas for $\phi$ only the intermediate layers
are important.

For these reasons there is a broad spectrum of theoretical possibilities
in the magnitude of the ratio $\phi/\alpha$. For a small correlation
length and near equipartition fields, one expects $\phi\ll \alpha$,
whereas for weaker fields and a larger correlation length the reverse
should be true, $\phi \gg \alpha$.

\paragraph*{One-temperature assumption}
We assumed here that the electron and proton temperatures are
equal. This is clearly so at the cold disk and the transition
layer. For the hot corona, the Coulomb interactions may be inefficient
in keeping the two temperatures equal. However, even under the
assumption that almost all the heating goes into protons, and only the
Coulomb interactions operate, the two temperatures in a NRAF are
approximately equal at radii $R\simgt 10^3 R_S$ (e.g. see Fig. 5 in
Narayan 2002). Thus in the most ``interesting'' for us range of radii
the one-temperature assumption should be quite reasonable.

In the innermost regions of the flow the protons may be much hotter
than the electrons. Clearly the heat flux is then dominated by the
protons. However, this heat flux should be of order $\phi \rho c_s^3$,
i.e. only slightly lower. Hence our main argument remains the same:
saturated thermal conduction will induce condensation of the hot flow
when the viscous heating is weak. The physics of cooling in the
transition layer will however be different for small radii
(e.g. Spruit \& Haardt 2000).


\begin{thebibliography}{}

\bibitem[]{} Abramowicz, M.A., Chen, X., Kato, S., Lasota, J.-P., \&
Regev, O. 1995, ApJL, 438, L37

\bibitem[]{} Albritton, J.R., Williams, E.A., Bernstein, I.B., \&
Swartz, K.P. 1986, PhRvL, 57, 1887

\bibitem[]{} Baganoff, F.K., Maeda, Y., Morris, M. et al. 2003, ApJ,
591, 891

\bibitem[]{} Balbus, S.A. 1986, ApJ, 304, 787

\bibitem[]{} Balbus, S.A., \& Hawley, J.F. 1991, ApJ, 376, 214

\bibitem[]{} Blandford, R., \& Begelman, M.C. 1999, MNRAS, 303, L1


\bibitem[]{} Cannizzo, J.K. 1998, in ``Wild Stars in the Old West'',
Eds. S. Howell, E. Kuulkers, \& C. Woodward (San Francisco: ASP),
p. 308

\bibitem[]{} Cowie, L.L., \& McKee, C.F. 1977, ApJ, 211, 135

\bibitem[]{} Dewangan, G.C., Griffiths, R.E., Di Matteo, T., \&
Schurch, N.J. 2004, to appear in ApJ (astro-ph/0402327).

\bibitem[]{} Ditmire, T., Gumbrell, E.T., Smith, R.A., Djaoui, A., \&
Hutchinson, M.H.R. 1998, PhRvL. 80, 720



\bibitem[]{} Dullemond, K. 1999, A\&A, 341, 936

\bibitem[]{} Elvis, M., et al. 1994, ApJS, 95, 1

\bibitem[]{} Eppelrein, E.M., \& Short, R.W. 1991, Phys. Fluids B
3(11), 3092

\bibitem[]{} Esin, A. 1997, ApJ, 482, 400


\bibitem[]{} Frank, J., King, A., \& Raine, D. 1992, Accretion Power
in Astrophysics (Cambridge, UK: Cambridge University Press)


\bibitem[]{} Ghez, A., et al. 2003, ApJL, 586, 127

\bibitem[]{} Gierlinski, M. et al. 1999, MNRAS, 309, 496


\bibitem[]{} Ho, L.C. 1999, ApJ, 516, 672

\bibitem[]{} Ho, L.C., 2003, Active Galactic Nuclei: from Central
Engine to Host Galaxy, ed. S. Collin, F. Combes, \& I. Shlosman (San
Francisco: ASP)




\bibitem[]{} Di Matteo, T., Allen, S.W., Fabian, A.C., Wilson, A.S.,
\& Young, A.J. 2003, ApJ, 582, 133


\bibitem[]{} Field, G.B. 1965, ApJ, 142, 531

\bibitem[]{} Kolykhalov, P.I., \& Sunyaev, R.A. 1980,
Sov. Astron. Lett., 6, 357

\bibitem[]{} Krolik, J.H., McKee, C.F., \& Tarter, C.B. 1981, ApJ,
249, 422


\bibitem[]{} Lasota, J.-P. 2001, New Astronomy Reviews, 45, 449


\bibitem[]{} Liu, B.F., Meyer, F., \& Meyer-Hofmeister, E. 1997,
A\&A, 328, 247

\bibitem[]{} Luciani, J.F., Mora, P., \& Virmont, J. 1983,
PhRvL 51, 18

\bibitem[]{} Luciani, J.F., Mora, P., \& Pellat, R. 1985, Phys. Fluids
28(3), 835

\bibitem[]{} McClintock, J.E., Narayan, R., et al. 2003, to appear in
ApJ (astro-ph/0304535)

\bibitem[]{} McKee, C.F., \& Cowie, L.L.  1977, ApJ, 215, 213

\bibitem[]{} McKee, C.F., \& Begelman, M.C. 1990, ApJ, 358, 392

\bibitem[]{} Meyer, F., \& Meyer-Hofmeister, E. 1994, A\&A, 288, 175
(MMH94)

\bibitem[]{} Meyer-Hofmeister, E., \& Meyer, F. 2001, A\&A, 380, 739


\bibitem[]{} Narayan, R., \& Yi, I. 1994, ApJL, 428, L13

\bibitem[]{} Narayan, R., \& Yi, I. 1995, ApJ, 452, 710

\bibitem[]{} Narayan R. 2002, in Lighthouses of the Universe,
Gilfanov, M., Sunyaev, R., \& Churazov, E. (editors), Springer


\bibitem[]{} Nayakshin, S., \& Svensson, R. 2001, ApJ, 551, L67

\bibitem[]{} Nayakshin, S., Cuadra, J., \& Sunyaev, R. 2004, A\&A,
415, 175

\bibitem[]{} Nayakshin, S., \& Sunyaev, R. 2004, in preparation.

\bibitem[]{} Parker, E.N. 1973, Interplanetary Dynamical Processes,
New York, Interscience.

\bibitem[]{} Penston, M.V., \& Brown, F.E. 1970, MNRAS, 150, 373

\bibitem[]{} Ptak, A., Terashima, Y., Ho, L.C., \& Quataert, L. 2004,
to appear in ApJ (astro-ph/0401525)

\bibitem[]{} Proga, D., \& Begelman, M.C., 2003, ApJ, 582, 69

\bibitem []{} Raymond, J.C., Cox, D.P., Smith, B.W. 1976, ApJ, 204,
290

\bibitem[]{} \rozanska, A., \& Czerny, B. 2000a, A\&A, 360, 1170
\bibitem[]{} \rozanska, A., \& Czerny, B. 2000b, MNRAS, 316, 473

\bibitem[]{} Schurtz, G.P., Nicolai, Ph. D., \& Busquet, M. 2000,
Phys. of Plasmas, 7, 4238

\bibitem[]{} Shakura, N.I., \& Sunyaev, R.A. 1973, A\&A, 24, 337

\bibitem[]{} Siemiginowska, A., Czerny, B., \& Kostyunin, V. 1996,
ApJ, 458, 491

\bibitem[]{} Spitzer, L., \& Harm, R. 1953, Phys. Rev.,
89, 977

\bibitem[]{} Spitzer, L. 1962, ``Physics of fully ionized gases'', 
New York, Interscience.

\bibitem[]{} Spruit, H.C., \& Haardt, F. 2000, MNRAS, 315, 751



\bibitem[]{} Wood, J., Horne, K., Berriman, G., Wade, R., O'Donoghue,
D., Warner, B. 1986, MNRAS, 219, 629


\bibitem[]{} Zel'dovich, Ya.B., \& Pickel'ner, S.B. 1969, JETP 29, 170

\end{thebibliography}
\end{document}